\documentclass[11pt,a4paper]{article}
\pdfoutput=1
\usepackage{jcappub}
\usepackage[usenames,dvipsnames]{xcolor}
\usepackage{graphicx}
\usepackage{amsmath,amssymb,mathrsfs}

\newcommand{\be}{\begin{equation}}
\newcommand{\ee}{\end{equation}}
\newcommand{\bea}{\begin{eqnarray}}
\newcommand{\eea}{\end{eqnarray}}

\newcommand{\eqn}{\begin{eqnarray}}
\newcommand{\eqnx}{\end{eqnarray}}

\begin{document}

\title{BPS Skyrme neutron stars in generalized gravity}
%\author{authors}
%\affiliation{affiliation}

\author[a]{C. Adam}
\emailAdd{adam@fpaxp1.usc.es}
\author[b]{M. Huidobro}
\emailAdd{miguel.huidobro.garcia@usc.es}
\author[c]{R. Vazquez}
\emailAdd{vazquez@fpaxp1.usc.es}
\author[d]{A. Wereszczynski}
\emailAdd{andrzej.wereszczynski@uj.edu.pl}

\affiliation[a,b,c]{Departamento de F\'isica de Part\'iculas, Universidad de Santiago de Compostela and Instituto Galego de F\'isica de Altas Enerxias (IGFAE) E-15782 Santiago de Compostela, Spain}

\affiliation[d]{Institute of Physics,  Jagiellonian University, Lojasiewicza 11, Krak\'{o}w, Poland}

\abstract{ We study the coupling of nuclear matter described by the BPS Skyrme model to generalized gravity. Concretely, we consider the Starobinsky model which provides the leading-order correction to the Einstein-Hilbert action.
Static solutions describing neutron stars are found both for the full field theory and for the mean-field approximation. We always consider the full Starobinsky model in the nonperturbative approach, using appropriately generalized shooting methods for the numerical neutron star calculations. Many of our results are similar to previous investigations of neutron stars for the Starobinsky model using other models of nuclear matter, but there are some surprising discrepancies. The "Newtonian mass" relevant for the surface redshift, e.g., results larger than the ADM mass in our model, in contrast to other investigations. This difference is related to the particularly high stiffness of nuclear matter described by the BPS Skyrme model and offers an interesting possibility to distinguish different models of nuclear matter within generalized gravity. } 

%\begin{abstract}
%\end{abstract}
\maketitle 
%Índice
%\newpage{}
%\begin{abstract} 
%\end{abstract}
%\thispagestyle{empty}
%\tableofcontents

\section{Introduction}
Sufficiently massive stars end up either as black holes or as neutron stars \cite{Glendenning:1997wn,lattimer2004physics,Haensel:2007yy}, in general. 
Neutron stars are, therefore, the remnants of many of the most massive stars in our universe. They have typical radii of roughly 10 km and masses around the mass of the Sun, hence they provide interesting scenarios in which General Relativity (GR) plays an important role. Besides, neutron stars are essentially composed of neutrons at very high densities. The study of these stars, therefore, also allows to extract information about nuclear matter and, in particular, about its equation of state (EoS), $\rho (p)$, which relates the pressure $p$ and the energy density $\rho$, at densities that are still not accessible in man-made laboratories.

Neutron stars are affected by intense gravitational fields (more precisely, by high curvature effects) more strongly than any other currently observable physical system in the universe. They are, therefore, perfect natural laboratories to investigate the consequences of this high curvature. In particular, they allow us to study deviations from General Relativity and to constrain the free parameters in theories describing these deviations, or even to discard them by comparing their predictions with the observed data. Such deviations, or Extended Theories of Gravity (ETG), are completely natural from an effective field theory point of view, where quantum gravity corrections should induce further terms in the low-energy effective action of the gravitational field, in addition to the Einstein-Hilbert (EH) action.
The \textit{f(R)} theories \cite{sotiriou2010f,de2010f,nojiri2007introduction,nojiri2011unified,capozziello2011extended,Nojiri:2017ncd} are a specific class of these ETG in which the EH action linear in the Ricci scalar $R$ is replaced by a generic function $f(R)$. These theories are motivated, first, by their relative simplicity when it comes to solve the modified Einstein equations and, second, from a cosmological point of view, by the possibility to explain the acceleration of the universe (the dark energy problem) or the abundance of non-baryonic matter in the galaxies (the dark matter problem) thanks to the presence of an additional degree of freedom, which is usually related to a scalar field known as the scalaron. Besides, these $f(R)$ theories lead to modified Einstein equations which are of more than second order, but they avoid the Ostrogradski instability \cite{woodard2007avoiding,ostrogradsky1850memoires}.

In particular, we shall consider the theory 
\begin{equation}
f(R) = R - \alpha R^2
\end{equation}
 (the minus sign in front of the quadratic term is because of our metric convention (\ref{metric})), also known as the Starobinsky model \cite{starobinsky1996new}. The reason for this choice is that standard GR is in excellent agreement with all current astrophysical and cosmological observations. Any extension of GR, therefore, should approach GR in the limit of small curvature. But the most natural way to achieve this is by a power series expansion $f(R) = R + \sum_{i=2}^\infty c_iR^i$, and the term quadratic in $R$ is the leading-order correction to the EH action.

The usual procedure to solve neutron stars is the Tolman-Oppenheimer-Volkoff (TOV) \cite{oppenheimer1939massive, tolman1939static} approach, in which the Einstein equations are solved assuming a perfect-fluid stress-energy tensor. To close the set of equations, an EoS is needed which should result from a theory of nuclear matter (i.e., strongly interacting matter in the low-energy sector). As Quantum Chromodynamics (QCD) is non-perturbative at low energies, we currently are not able to use it to derive an EoS for nuclear matter. The EoS is obtained either from phenomenological models of nuclear physics which are at least partially determined by experimental fits, or from effective field theories (EFT) motivated by some arguments from the fundamental theory. The Skyrme model \cite{skyrme1994non, skyrme1962unified} is an interesting EFT, principally motivated by the \textit{large $N_c$ limit} \cite{tHooft:1973alw,Witten:1979kh} of QCD, in which baryons and nuclei are collective (solitonic) excitations of the fundamental field \cite{manton2004topological,Weigel:2008zz,Shnir:2018yzp,Ma:2019kiq}.  
The standard version of the Skyrme model (the model originally proposed in \cite{skyrme1962unified}) has been applied to the description of nucleons \cite{Adkins:1983ya} and some light nuclei (see, e.g., \cite{Braaten:1985np,Weigel:1986zc,Battye:2009ad}) with certain success. 
Moreover, in the last decade several generalizations have been introduced \cite{adam2010skyrme,Sutcliffe:2010et,Gillard:2015eia,Gudnason:2016tiz,Gudnason:2018jia,Naya:2018kyi,Adam2020diel} which lead to significant improvements of the shortcomings of the original Skyrme model. In particular, in this paper we will use an extension of the model known as the BPS Skyrme model \cite{adam2010skyrme,adam2010bps,naya2015bps,Adam2015-in-RhoZahed}. The BPS Skyrme model has BPS solutions, \textit{i.e.}, the mass of its solutions is quantized by its topological charge (which is identified with the baryonic number).  Therefore, the model is capable of describing the small binding energies of  physical nuclei \cite{Adam:2013wya}. Further, when considered as a submodel within the class of generalized Skyrme models, the BPS Skyrme model provides the leading contribution to the nuclear EoS at large densities \cite{Adam:2015lra}, so its inclusion is unavoidable for a realistic description of nuclear matter at high density.
In addition, it implies another interesting property: the TOV procedure requires a perfect-fluid stress-energy tensor but, usually, an EFT does not describe a perfect fluid, hence to relate the pressure and the energy density, a macroscopic (mean-field) approximation must be used. This mean-field approximation defines, by construction, a spatially constant energy density and, therefore, a barotropic EoS. Gravity, however, couples to the derivatives of the fields, so any deviation from spatial constancy can produce effects in the observables. The BPS Skyrme model stress-energy tensor has the form of a perfect fluid \cite{adam2015bps, adam2015neutron}, so we can couple the microscopic (exact) degrees of freedom to gravity or consider a mean-field approximation, and compare the resulting effects. 

This paper is organised as follows: In the second section we write down the modified Einstein equations and the resulting system of ordinary differential equations (ODE) when spherical symmetry is imposed. Further, we discuss the interpretation of the new degree of freedom in the \textit{f(R)} theories. In the third section we briefly discuss the relevant aspects of the BPS Skyrme model. In section 4, we perform a numerical analysis to solve the equations. To do that, a double shooting method is required to find the correct initial conditions. In the fifth section we present the results of our integration, and in the last section we discuss these results and our conclusions. Our calculations are performed in the Jordan frame with the metric formalism. We choose units such that the speed of light $c=1$. For lengths and masses we use either astrophysical units (km and solar masses $M_\odot$) or nuclear physics units (fm and MeV).

\section{Generalized Einstein equations}
The \textit{f(R)} theories consist in the following modification of the Einstein-Hilbert action,
\begin{equation}
    S = \frac{1}{2\kappa} \int \: d^4x \: \sqrt{-g}\: f(R) + S_{\text{matter}},
    \label{fRaction}
\end{equation}

where $\kappa = 8\pi G/c^4$, $g$ denotes the determinant of the metric and $f(R)$ is a generic function of the Ricci scalar $R$. General Relativity can be recovered by setting $f(R)=R$.

The new Einstein equations may be obtained by varying this action with respect to the metric following \cite{blau2011lecture}.

%
%To do that is important to know the next relations:
%\begin{align}
 %    &  \delta f = f_R \delta R\\
 %    &  \delta R = R_{\mu\nu}\delta g^{\mu\nu} + \delta R_{\mu\nu} g^{\mu\nu}\\
%     &  \delta R_{\mu\nu} = \nabla_{\lambda}\delta \Gamma^{\lambda}_{\mu\nu} - \nabla_{\nu}\delta \Gamma^{\lambda}_{\mu\lambda}\\
%     &  \delta \Gamma^{\rho}_{\mu\nu} = \frac{1}{2}g^{\rho\lambda}\left( \nabla_{\mu}\delta g_{\nu\lambda}+\nabla_{\nu}\delta g_{\mu\lambda}-\nabla_{\lambda}\delta g_{\mu\nu} \right),
%\end{align}
%
Then the modified Einstein equations are
\begin{equation}
    f_R R_{\mu\nu} - \frac{1}{2}g_{\mu\nu}f - \left(\nabla_{\mu}\nabla_{\nu} - g_{\mu\nu}\nabla^{\alpha}\nabla_{\alpha}  \right)f_R = \kappa T_{\mu\nu}
    \label{Einseq}
\end{equation}
where $f_R$ denotes the derivative of $f$ with respect to $R$.

Now, to obtain a system of differential equations we have to write an explicit form of the metric. As we are modeling static neutron stars, we will choose the spherically symmetric ansatz with the following signature,

\begin{equation}
    ds^2 = A\: dt^2 - B\:dr^2 - r^2\left(d\theta^2 + \sin^2\theta \: d\varphi^2 \right).
		\label{metric}
\end{equation}

The first difference with respect to General Relativity is that now the Ricci scalar is not fixed (algebraically)  just by the value of the pressure and the energy density. Instead, tracing the equation (\ref{Einseq}) we obtain a second order differential equation for $R$ that we have to solve. The key point is that we now have an additional degree of freedom in comparison to the General Relativity case, precisely provided by the $f_R(R)$ term. The equations for the metric components $A$ and $B$ can be obtained from the $tt$ and $rr$ components, respectively. We also need an equation to describe the pressure along the radius of the star, which is easily obtained from the conservation of the stress-energy tensor: $\nabla_{\mu} T^{\mu\nu} = 0$. The resulting system of ordinary differential equations is ($B' \equiv (d/dr)B$, etc.) 
\begin{align}
     &  B' = \frac{2rB}{3f_R}\left[ \frac{\kappa}{2}B(\rho + 3p) - Bf - f_R \left( -\frac{BR}{2} + \frac{3A'}{2Ar} \right) - \left( \frac{3A'}{2A} + \frac{3}{r} \right)f_{2R}R' \right], \label{eqB} \\[3mm]
	&  p' = - \frac{\rho + p}{2} \frac{A'}{A}, \label{eqp} \\[3mm]
	&  A'' = \frac{A'B'}{2B} + \frac{A'^2}{2A} + \frac{2B'A}{rB} + \frac{2A}{f_R}\left(-\frac{\kappa}{2}Bp + \frac{Bf}{2} + \left(\frac{A'}{2A} + \frac{2}{r} \right)f_{2R}R' \right), \label{eqA} \\[3mm]
	&  R'' = -\frac{f_{3R}}{f_{2R}}R'^2 - \frac{B}{3f_{2R}}\left( \frac{\kappa}{2}(\rho - 3p) + 2f - f_R R \right) + \left( -\frac{A'}{2A} + \frac{B'}{2B} - \frac{2}{r} \right)R' \label{eqR}.
\end{align}
This is a system of 4 equations for the 5 unknowns $A$, $B$, $R$, $\rho$ and $p$ and, therefore, one additional equation is required to close the system. Further, $f_{2R}(R) \equiv (\partial^2/\partial R^2) f$, etc.

\subsection{Correspondence between \textit{f(R)} gravity and scalar-tensor theories}
Next, we will briefly comment on the relation between \textit{f(R)} theories and scalar-tensor theories. They are, in fact, completely equivalent. As explained, we will focus on the theory $f(R) = R - \alpha R^2$, which has the important property that $f_{2R}(R)  \neq 0$.

The Brans-Dicke action \cite{brans1961mach} takes the form
\begin{equation}
    S_{BD} = \frac{1}{2\kappa} \int \:d^4x \: \sqrt{-g}\:\left[ \phi R - \frac{\omega}{\phi}g^{\mu\nu}\nabla_{\mu}\phi\nabla_{\nu}\phi - V(\phi) \right] + S_{\text{matter}}.
    \label{BDaction}
\end{equation}
To see the correspondence, we can rewrite the action (\ref{fRaction}) in a new dynamically equivalent form with a new scalar field $\chi$,
\begin{equation}
		S = \frac{1}{2\kappa} \int \: d^4x \: \sqrt{-g} \: \left( f\left( \chi \right) + f'\left( \chi \right)\left( R - \chi \right) \right).
\end{equation}
From the scalar field equation we find that $R = \chi$ iff $f_{2R}(\chi) \neq 0$, then redefining the scalar field as $\phi := f'(\chi)$ and $V(\phi) = \chi(\phi)\phi - f(\phi)$ we arive at the Brans-Dicke action for $\omega = 0$.

From the action (\ref{BDaction}) we can define the mass of the scalar field $\phi$, by obtaining its equation of motion and identifying the terms with those of the Klein-Gordon equation. The Einstein equations and the scalar field equation of this action are
\begin{align}
    \phi G_{\mu\nu} +\frac{\omega}{2\phi}g^{\mu\nu}\nabla_{\alpha}\phi\nabla^{\alpha}\phi +\frac{1}{2}g_{\mu\nu}&V(\phi) - \frac{\omega}{\phi}\nabla_{\mu}\phi \nabla_{\nu}\phi -(\nabla_{\mu}\nabla_{\nu} - g_{\mu\nu}\nabla_{\alpha}\nabla^{\alpha})\phi = \kappa T_{\mu\nu} ,\\[2mm]
    &  \Box \phi - \frac{1}{3+2\omega}\left( \kappa T + \phi V'(\phi) - 2V(\phi) \right) = 0. \label{eqmotionphi}
\end{align}

The equation of motion of the field can be expressed as a Klein-Gordon equation, defining an effective potential \cite{capone2010jumping,faraoni2009scalar}, $\frac{dV_{\text{eff}}}{d\phi} = -\frac{1}{3+2\omega}\left( \phi V' - 2V \right)$. Then the mass is related to the second derivative of this effective potential, and equation (\ref{eqmotionphi}) admits the usual Yukawa-like solution $\phi(r) \propto \exp\left( -m(\phi)r \right)/r$ with $m$ defined as explained above. Now we can obtain the mass of the new degree of freedom in our theory in terms of $f(R)$ and its derivatives. Setting $\omega = 0$ and the potential given above we have for a generic $f(R)$ theory \cite{faraoni2009scalar}
\begin{equation}
    m^2 = -\frac{1}{3}\frac{f_R-Rf_{2R}}{f_{2R}}.
\end{equation}
In our case $f(R) = R - \alpha R^2$, we obtain that the mass of the field is $m^2 = \frac{1}{6\alpha}$, thus the sign of $\alpha$ must be positive in order to have a real scalar field.

\section{The BPS Skyrme model and its EoS}
As explained, we need a further relation to close the system of equations (\ref{eqB}) - (\ref{eqR}), and this relation should be derived from a model for nuclear matter relevant for the conditions inside a neutron star.
Concretely, we shall use the BPS Skyrme model as our model of nuclear matter, which provides us with two possibilities to relate the pressure and the energy density. We may either derive their exact expressions in terms of the (microscopic) Skyrme field variables from the stress-energy tensor, thus coupling the microscopic degrees of freedom to gravity, or we may perform a mean-field approximation to obtain a barotropic EoS.

\subsection{Exact case}
The BPS Skyrme model in flat (Minkowski) space is given by the lagrangian \cite{adam2010skyrme} 
\begin{align}
    &  \mathcal{L}_{BPS} = \mathcal{L}_0 + \mathcal{L}_6 = -\mu^2 \mathcal{U}(U) - \lambda^2 \pi^4 \mathcal{B}_\nu\mathcal{B}^\nu, \label{lagrangian} \\[2mm]
    &  L_{\mu} = U^{\dagger}\partial_{\mu}U, \hspace{3mm} U = e^{i\xi \vec{n}\vec{\tau}} = \cos\xi\: I + \sin\xi \:\vec{n}\vec{\tau} \in \text{SU(2)}, \label{Skfield} \\
&    \mathcal{B}^{\mu} = \frac{1}{24\pi^2}\epsilon^{\mu\nu\alpha\beta}L_{\nu}L_{\alpha}L_{\beta} , \hspace{3mm} N = \int d^3x \: \mathcal{B}^0,
\end{align}
where $\mu^2$ and $\lambda^2$ are constants fixed by nuclear experimental data and $\mathcal{U}$ is a potential that will be specified in the next sections. We will, however, always assume that the potential only depends on the trace of $U$, i.e., on the Skyrme profile function $\xi$, $\mathcal{U} = \mathcal{U}(\xi)$, such that it breaks chiral symmetry but respects isospin. Further, $\mathcal{B}^{\mu}$ is the topological current of the model which is defined in terms of the fundamental SU(2) field $U$ of the theory, and from which the baryon number $N$ can be calculated as a topologically conserved charge
(we use $N$ to denote the baryon number to avoid confusion with the metric component $B$).

One of the interesting properties of the BPS Skyrme model, and the reason of its name, is that we can find BPS solutions that saturate a Bogomol'nyi bound \cite{Bogomolny:1975de}. This bound can be extracted from the static energy functional, and it leads to a BPS equation which can be solved analytically \cite{adam2010skyrme} and yields solutions whose energy is exactly proportional to the baryonic number, which is a welcome property which physical nuclei obey approximately. To derive this equation we just have to reconstruct the Bogomol'nyi bound in the energy functional,
\begin{align}
	\notag &E = \int d^3x \: \left( \lambda^2 \pi^4 \mathcal{B}^2_{0} + \mu^2 \mathcal{U} \right) = \\[2mm]
	\int& d^3x \: \left( \lambda \pi^2\mathcal{B}_0 \pm \mu \sqrt{\mathcal{U}} \right)^2 
	\mp 2\pi^2\lambda\mu \int d^3x \: \mathcal{B}_0 \sqrt{\mathcal{U}} \\[2mm] &
	\geq 2\pi^2\lambda\mu \left| \int d^3x \: \mathcal{B}_0 \sqrt{\mathcal{U}} \right| = \lambda \mu |N| \int \Omega_{\mathbb{S}^3} \sqrt{\mathcal{U}},
\end{align}
where the last integral is over field space SU(2) which, as a manifold, is just the unit three-sphere $\mathbb{S}^3$ (here, $\Omega_{\mathbb{S}^3}$ is the volume form on $\mathbb{S}^3$).  The inequality turns into an equality if the BPS equation
\begin{equation}
\lambda \pi^2\mathcal{B}_0 = \pm \mu \sqrt{\mathcal{U}} \label{BPSeq}
\end{equation}
holds.

The stress-energy tensor is obtained by varying the BPS action in curved space,
\begin{equation} \label{bps-act}
S_{\rm BPS} = \int d^4 x |g|^\frac{1}{2} \left( -\lambda^2 \pi^4 |g|^{-1} g_{\alpha\beta} {\cal B}^\alpha {\cal B}^\beta - \mu^2 {\cal U} \right) ,
\end{equation}
w.r.t. the metric [$T^{\mu\nu} = -(2/\sqrt{|g|})(\delta / \delta g_{\mu\nu}) S_{\rm BPS}$],
and we can identify the energy density and the pressure by comparing with a perfect fluid,
\begin{align}
    &  T^{\mu\nu} = \frac{2\lambda^2\pi^4}{|g|}\mathcal{B}^{\mu}\mathcal{B}^{\nu} - \left(\frac{\lambda^2\pi^4}{|g|}g_{\alpha\beta}\mathcal{B}^{\alpha}\mathcal{B}^{\beta} - \mu^2 \mathcal{U} \right)g^{\mu\nu}, \\[2mm]
    &  \rho = \frac{\lambda^2\pi^4}{|g|}g_{\alpha\beta}\mathcal{B}^{\alpha}\mathcal{B}^{\beta} + \mu^2 \mathcal{U}, \\[2mm]
	&  p = \frac{\lambda^2\pi^4}{|g|}g_{\alpha\beta}\mathcal{B}^{\alpha}\mathcal{B}^{\beta} - \mu^2 \mathcal{U} \label{pdef}.
\end{align}

From this definitions we can extract the exact, off-shell (i.e., solution-independent) relation for the BPS Skyrme model relating $\rho$, $p$ and the Skyrme profile field $\xi$,
\begin{equation} \label{exact-rel}
    \rho = p + 2\mu^2 \mathcal{U}.
\end{equation}

In order to use these exact expressions for a neutron star calculation, we now need an ansatz for the Skyrme field (\ref{Skfield}) which leads to a static and spherically symmetric energy density and pressure.
Concretely, we choose the axially symmetric (generalized hedgehog) ansatz 
\begin{equation} \label{axisym-ansatz}
	\xi = \xi(r), \hspace{3mm} \vec{n} = (\sin\theta \cos N\phi,\sin\theta \sin N\phi,\cos\theta) ,
\end{equation}
which is compatible with the full field equations and the metric (\ref{metric}). It leads to
%For this choice we obtain the expression of $\mathcal{B}^{\mu}$:
\begin{equation}
	\mathcal{B}^{0} = -\frac{N}{2\pi^2}\xi' (r)\sin^2\xi(r)\sin\theta, \hspace{3mm} \mathcal{B}^{1} = \mathcal{B}^{2} = \mathcal{B}^{3} = 0, \label{Bmu}
\end{equation}
\begin{equation} \label{p(r)}
p(r)= \frac{\lambda^2N^2}{4}\, \frac{\xi'^2(r) \sin^4 \xi (r)}{r^4 B(r)} -\mu^2 \mathcal{U}.
\end{equation} 
In principle, we could express both $p$ and $\rho$ as functions of $\xi$ and its derivative via  (\ref{p(r)}) and (\ref{exact-rel}) and insert them into the system of equations (\ref{eqB}) - (\ref{eqR}) which then closes by itself. 

For our numerical calculations it is, however, more useful to keep both $p$ and $h$ as independent field variables (where we define the new variable $h := \sin^2 (\xi/2)$, for convenience) and to add the new equation 
\begin{equation}
    h' = \frac{\sqrt{B} r^2}{2N\lambda} \frac{1}{\sqrt{h (1-h)}} \left( p + \mu^2 \mathcal{U}(h) \right)^{1/2} \label{eqh}
\end{equation}
to the system, which easily follows from (\ref{p(r)}).

One important difference with respect to the mean-field case is that the exact calculation with the generalized hedgehog ansatz (\ref{axisym-ansatz}) introduces the baryon number $N$ as an input parameter. When we solve the system numerically, therefore, the value of the pressure in the center of the star is not an input parameter. Instead, we have to find the correct value of $p_0 \equiv p(r=0)$ for a given baryon number via a shooting method.

\subsection{Mean-Field limit}

For the mean-field limit, we need expressions for the average energy density $\bar \rho$ and the average baryon density $\bar n$ (in flat space, because they only contain information about the nuclear interactions). Their construction starts from the observation that for a perfect fluid like the BPS Skyrme model in flat Minkowski space, the pressure (\ref{pdef}) is constant for all static solutions as a consequence of energy-momentum conservation,
\begin{equation}
\partial_\mu T^{\mu\nu} = \partial_i T^{ij} = \delta^{ij}\partial_i p =0 \quad \Rightarrow \quad 
p \equiv \lambda^2\pi^4 (\mathcal{B}^0)^2 - \mu^2 \mathcal{U} = {\rm const.}
\end{equation} 
The equation $p=$ const. is a first integral of the static field equations, and $p$ is the integration constant. The BPS equation (\ref{BPSeq}) corresponds to the special case $p=0$. 
Further, it turns out that, as a consequence of the symmetries of the model, both the energy and the volume are the {\em same} for all solutions with the same pressure and may be re-expressed as target space integrals \cite{adam2015neutron}. Concretely,
\begin{align}
    &  E(p) = 2\pi \lambda \mu N \widetilde{E}(p), \\[2mm]
    &  \widetilde{E}(p) = \int_0^\pi \: d\xi \: \sin^2\xi \: \frac{2\mathcal{U}+p/\mu^2}{\sqrt{\mathcal{U}+p/\mu^2}}
\end{align}
and
\begin{align}
    &  V(p) = 2\pi N \frac{\lambda}{\mu}\widetilde{V}(p), \\[2mm]
    &  \widetilde{V}(p) = \int_0^\pi \: d\xi \: \sin^2\xi \: \frac{1}{\sqrt{\mathcal{U}+p/\mu^2}},
\end{align}
which permits us to define the average energy density and, therefore, the required mean-field EoS
\begin{equation}
    \bar{\rho}(p) = \frac{E(p)}{V(p)}.
\end{equation}
In particular, the average energy density does not depend on the baryon number, as expected. 
This implies that, in contrast to the exact case, 
 now the pressure in the center is an input parameter, whereas the baryon number is a derived quantity.
To obtain it in the mean-field case, we can calculate the integral of the conserved baryon current $J_N^{\mu}$, which can be expressed in terms of the proper baryon number density $n$ \cite{Weinberg:1972kfs},
\begin{align}
    &  n = g_{\mu\nu}u^{\mu}J_N^{\nu} = \sqrt{A}J_N^0, \\[2mm]
    N = 4\pi  \: &\int_0^{R_s} \: \sqrt{AB}\: J_N^0 \:r^2 dr = 4\pi  \: \int_0^{R_s} \sqrt{B} \: n \: r^2 dr.
\end{align}
Now the average baryon number density can be obtained as the number of baryons divided by the volume they occupy
\begin{equation}
    \bar n = \frac{N}{V} = \frac{\mu}{2\pi \lambda \widetilde{V}}.
\end{equation}

We already have the tools to construct neutron stars in our $f(R)$ theory for the BPS Skyrme model, but we still have the freedom to choose a potential. We will perform our calculations for the potentials given below, where we also provide the numerical values of the model parameters, which have been determined by fitting to the binding energy per nucleon of infinite nuclear matter, $E_B = 16.3$ MeV, and to the nuclear saturation density, $n_0 = 1/V = 0.153 \:\text{fm}^{-3}$.
\begin{itemize}
    \item Step function potential: $\mathcal{U}=\Theta(h)$:
    This potential is not really phenomenologically motivated, but it is interesting because it reproduces a barotropic EoS in both cases, then the exact case coincides with its mean-field limit.
    \begin{equation*}
        \mu^2 = 70.61\: \text{MeV/$\text{fm}^3$}, \hspace{2mm} \lambda^2 =30.99\: \text{MeV $\text{fm}^3$} .
    \end{equation*}
    %\begin{equation}
        %\bar{\rho} = p + 2\mu^2.
    %\end{equation}
    \item Skyrme potential: $\mathcal{U} = 2h$:
    This is the usual potential used in the Skyrme model because it provides a mass to the pionic field. In this case the exact equation of state is no more spatially constant so it will not coincide with its mean-field version.
    \begin{equation*}
        \mu^2 =88.26\: \text{MeV/$\text{fm}^3$} , \hspace{2mm} \lambda^2 =26.88\: \text{MeV $\text{fm}^3$} .
    \end{equation*}
    %\begin{equation}
        %\bar{\rho} = \frac{\mu^2}{5}\left[ 2 - 3\frac{p}{\mu^2} + \frac{6}{1 + \frac{p}{\mu^2}\left( 1 - \frac{K[\frac{2}{2+p/\mu^2}]}{E[\frac{2}{2+p/\mu^2}]} \right)}\right],
    %\end{equation}
    %where $K$ and $E$ are the complete elliptic integral of the first and second kind respectively.
    \item Quadratic Skyrme potential: $\mathcal{U} = 4h^2$:
		This is the simplest choice for a potential with a quartic approach to the vacuum, which is also a natural choice for the BPS model.
    \begin{equation*}
        \mu^2 =141.22\: \text{MeV/$\text{fm}^3$} , \hspace{2mm} \lambda^2 =15.493\: \text{MeV $\text{fm}^3$} .
    \end{equation*}
    %\begin{equation}
        %\bar{\rho} = \mu^2 \left(\frac{p}{\mu^2}+\frac{5}{2} \frac{_3F_2 \left[\left\lbrace \frac{1}{2},\frac{7}{4},\frac{9}{4} \right\rbrace, \left\lbrace \frac{5}{2},3 \right\rbrace, -\frac{4 \mu^2}{p} \right]}{_3F_2 \left[\left\lbrace \frac{1}{2},\frac{3}{4},\frac{5}{4} \right\rbrace, \left\lbrace \frac{3}{2},2 \right\rbrace, -\frac{4 \mu^2}{p} \right]}  \right).
    %\end{equation}
    \item Partially flat potential: $\mathcal{U} = \left\{ \begin{array}{lcc}
																												16 h^2(1-h)^2 &, & h \in [0, 1/2] \\
																												1 &, & h \in [1/2,1].
																												\end{array}
																												\right.$
    \begin{equation*}
        \mu^2 = 121.08 \: \text{MeV/$\text{fm}^3$}, \hspace{2mm} \lambda^2 = 23.60\: \text{MeV $\text{fm}^3$}.
    \end{equation*}
    %\begin{equation}
        %\bar{\rho} = \left\{ 
					%						\right.
    %\end{equation}
\end{itemize}

\section{Numerical resolution}

For our numerical calculations, we first need to know the boundary conditions of all the variables and also of their derivatives, and these conditions are obtained analysing the Einstein equations (\ref{Einseq}) and the system (\ref{eqB}) - (\ref{eqR}).
We expand the field variables close to the center ($r \sim 0$) in powers of $r$. First of all, the smoothness condition at the center implies that the odd coefficients are null. 
%This conclusion is also supported by a numerical integration towards the center. 
Therefore, we can express our variables as
\begin{align*}
    &  A = a_0 + a_1 r^2 + ... \\[2mm]
    &  B = b_0 + b_1 r^2 + ... \\[2mm]
    &  R = R_0 + R_1 r^2 + ... \\[2mm]
    &  p = p_0 + p_1 r^2 + ... \\[2mm]
    &  \rho = \rho_0 + \rho_1 r^2 + ...
\end{align*}
From these expressions, we directly see that all the first derivatives cancel in the center. Thus we need the values of the variables in the center, but, as we have two equations that are of second order, we  also need the values of $a_1$ and $R_1$.

From the $(\theta,\theta)$ component of (\ref{Einseq}) we immediately obtain $b_0 = 1$ (this condition can, of course, also be found from (\ref{eqB}) - (\ref{eqR}), but it is not so obvious in this case). The value of $a_0$ cannot be determined in this way, because the system (\ref{eqB}) - (\ref{eqR}) can be re-expressed in terms of $A'/A$.  $A$, therefore, is determined only up to a multiplicative factor, and $a_0$ can take an arbitrary value. However, we want to have the Minkowski spacetime in the limit $r \rightarrow \infty$, which is characterized by
\begin{equation}
    A \rightarrow 1, \hspace{3mm} B \rightarrow 1, \hspace{3mm} R \rightarrow 0.
    \label{Minkcond}
\end{equation}
That is to say, if we find a solution $A_{\rm sol}$ such that $\lim_{r\to \infty} A_{\rm sol}(r) = a_\infty$, then the correctly normalized solution is $A(r) = (A_{\rm sol}(r)/a_\infty )$ and the correct boundary value at $r=0$ is  $a_0 =A(0) =(A_{\rm sol}(0)/a_\infty)$.

In the mean-field case, the pressure in the center is an input parameter. This means that for each (not too large) value $p_0$ we will find a neutron star solution. In the exact case, on the other hand, the input parameter is the baryon number $N$, so for fixed $N$ there is only one correct value for the pressure in the center, which we have to determine with a shooting method, as follows. In the exact case, we have to solve eq. (\ref{eqh}) in addition to the system (\ref{eqB}) - (\ref{eqR}), and the boundary conditions for the Skyrme profile $h$ are
\begin{equation}
    h(0) = 1, \hspace{3mm} h(R_s) = 0,
\end{equation}
where $R_s$ is the neutron star radius (i.e., the radius where the Skyrme field takes its vacuum value, which defines the star surface). But, obviously, the pressure must vanish at the star surface, as well. The shooting method, therefore, consists in solving the system many times for different values of $p_0$, until one finds the correct value that satisfies the condition that determines the radius of the star,
\begin{equation}
    p(R_s) = 0.
\end{equation}
The value of $\rho_0$ is completely determined in both cases (by the EoS in the mean-field case, and by eq. (\ref{exact-rel}) in the exact case). 

Finally, we need the initial value $R_0$ of the Ricci scalar. Unlike in General Relativity, the Ricci scalar is now not determined algebraically but satisfies its own second-order differential equation and, therefore, we do not know its initial value. The way to solve this problem is, again, by a shooting method with the condition of the Minkowski spacetime at large distances (\ref{Minkcond}). In order to be able to satisfy this condition, we have to integrate the system up to large distances, in contrast to the GR case where we just have to integrate until the radius of the star.

The shooting to determine $R_0$ required by $f(R)$ gravitational theories has been solved already for cases which are similar to our mean-field case, i.e., with a barotropic EoS (see for example \cite{astashenok2017realistic,kase2019neutron}).  In the exact case, on the other hand, we have to solve a double shooting problem for the pressure and the Ricci scalar. To solve this problem, we perform several shootings for the pressure until a sufficient accuracy is reached, then we change the value of the Ricci scalar. Besides, when we change the value of the Ricci scalar we also constrain the range of values in the shooting of the pressure. Repeating this iteration we obtain the required solutions.

The constraint in the pressure in each iteration is really important when solving the system because, as we explained, the solutions of the scalar field are Yukawa-like, \textit{i.e.}, exponential functions. We will have both the positive and the negative (exp($\pm m r$)) solutions, and as we want a finite solution and the mass is a real value, the growing exponential must be cancelled, but this can be obtained only with a very accurate initial condition for $R$. Another interesting feature that supports this argument is found when we change the values of $\alpha$. When $\alpha$ grows we are deviating more from General Relativity, but the mass of the field decreases, and we find that it is easier to reach a good accuracy in our solutions.

Finally, as we have two second-order equations, we have to start the integration at a small but nonzero value of $r$ and, therefore, need the values of $a_1$ and $R_1$. To obtain them, we just have to insert the expansions of the variables in the equations and take the limit $r=0$,
\begin{align}
    A''(0) = 2 a_1 =&\frac{2}{9\left( 1 -2\alpha R_0 \right)}\left( \kappa(2\rho_0 + 3p_0) + \frac{R_0}{2}- \frac{3}{4}\alpha R_0^2  \right), \\[2mm]
    &  R''(0) = 2 R_1 =\frac{1}{18\alpha}\left( \kappa \left(\rho_0 - 3p_0  \right) + R_0 \right).
\end{align}

%%%%%%%%%%%%%%
\section{Results}
%%%%%%%%%%%%%%
We have solved the Einstein equations with a fourth order Runge-Kutta method, so we can extract now the observables of the neutron star (the mass and radius) from the solutions. We find some interesting differences with respect to the GR case, but before showing the figures we will explain how to calculate the mass and comment briefly about the stability of the stars.

Once we integrate the system for a given value of the pressure in the center, we extract the radius of the star as the point $R_s$ where the pressure is zero. Then we maintain the integration with $p =\rho = 0$. 

 %%  Figure 1
\begin{figure}%[H]
	%\centering
		\hspace{-0.0cm}\includegraphics[width=16cm, height=10cm]{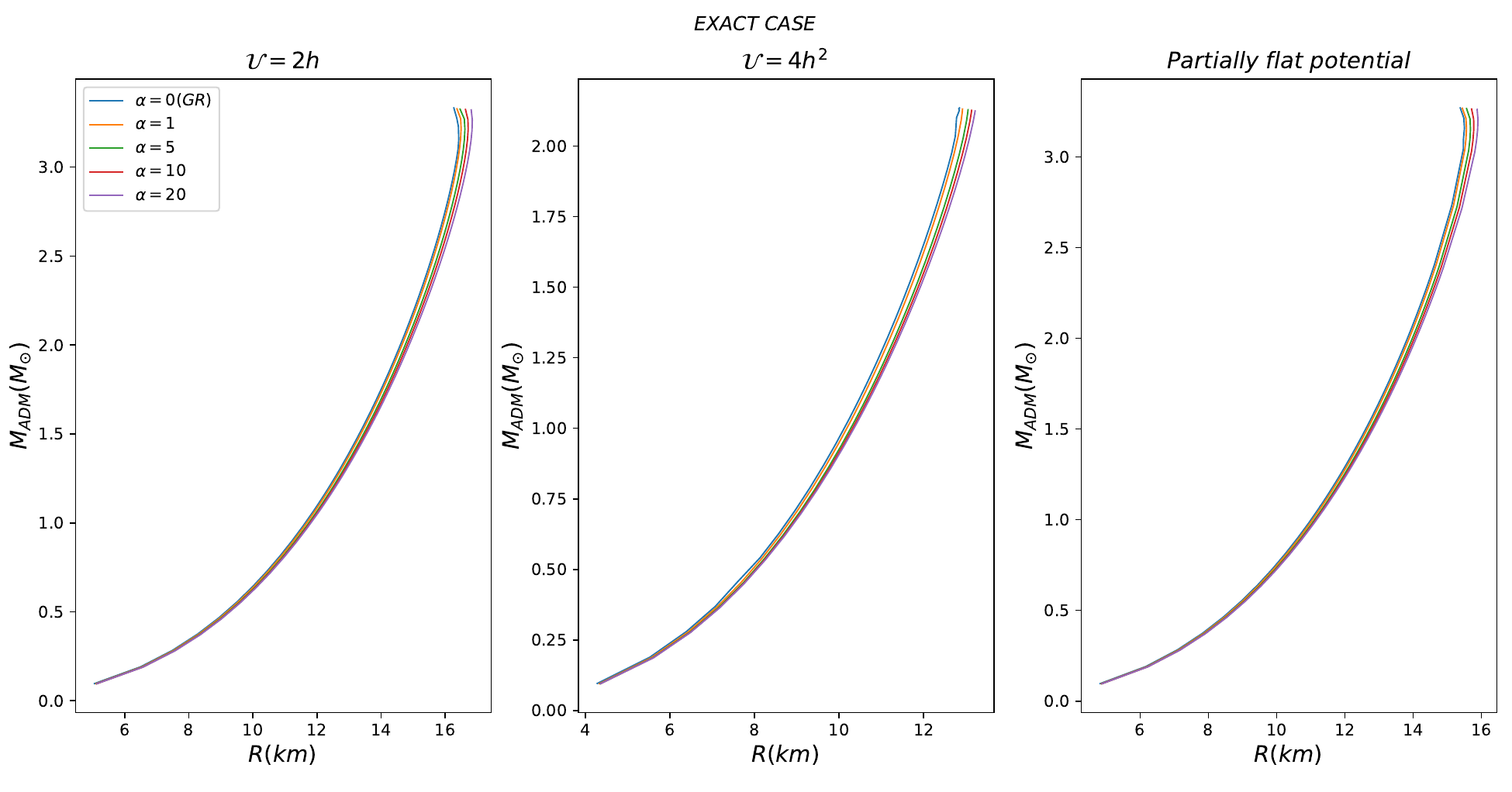}
		\caption{Mass-radius curve in the exact case, for three different potentials,  and for four different values of $\alpha$. The values of $\alpha$ are given in units of km$^2$. For a given ADM mass, the star radius always grows with $\alpha$. There is no appreciable difference in the maximum masses for different $\alpha$.}
	\label{fig:MvsR_BPS}
\end{figure}

In GR, the mass can be obtained by solving the differential equation
\begin{equation}
    \frac{dM_{\rho}}{dr} = \frac{\kappa}{2} r^2 \rho(r).
\end{equation}
This equation is obtained by taking the parametrization $B(r) = \left[ 1 - \kappa M(r)/(4\pi r) \right]^{-1}$ and by using the field equations of GR which imply $M(r)=M_\rho (r)$. In the region outside of the star, the space-time is described by the Schwarzschild metric (where $R = 0$ and $M=$ const.), thus we can identify $M_s \equiv M(R_s)$ with the mass of the star. 

However, in $f(R)$ gravity, we do not have the Schwarzschild solution for $r\ge R_s$, because $R$ satisfies its own differential equation and, in general, is nonzero for $r\ge R_s$, approaching zero only in the limit of large distances $r\to \infty$. As a consequence, the mass function 
\begin{equation}
M(r) = \frac{4\pi r}{\kappa} \left( 1 - B^{-1} (r) \right)
\end{equation}
 is no longer constant outside the star, and the surface mass $M_s =M(R_s)$ is different from the asymptotic or ADM (=Arnowitt-Deser-Misner) mass $M_\infty = \lim_{r\to \infty} M(r)$ as seen by a distant observer. $M_s$ is also different from $M_\rho (R_s)$, because it receives additional contributions from the curvature scalar inside the star radius (for a detailed discussion see \cite{Sbisa:2019mae}).

%% Figure 2
%\vspace{-1cm}
\begin{figure}%[H]
	%\centering
		\hspace{-0.0cm}\includegraphics[width=16cm, height=8cm]{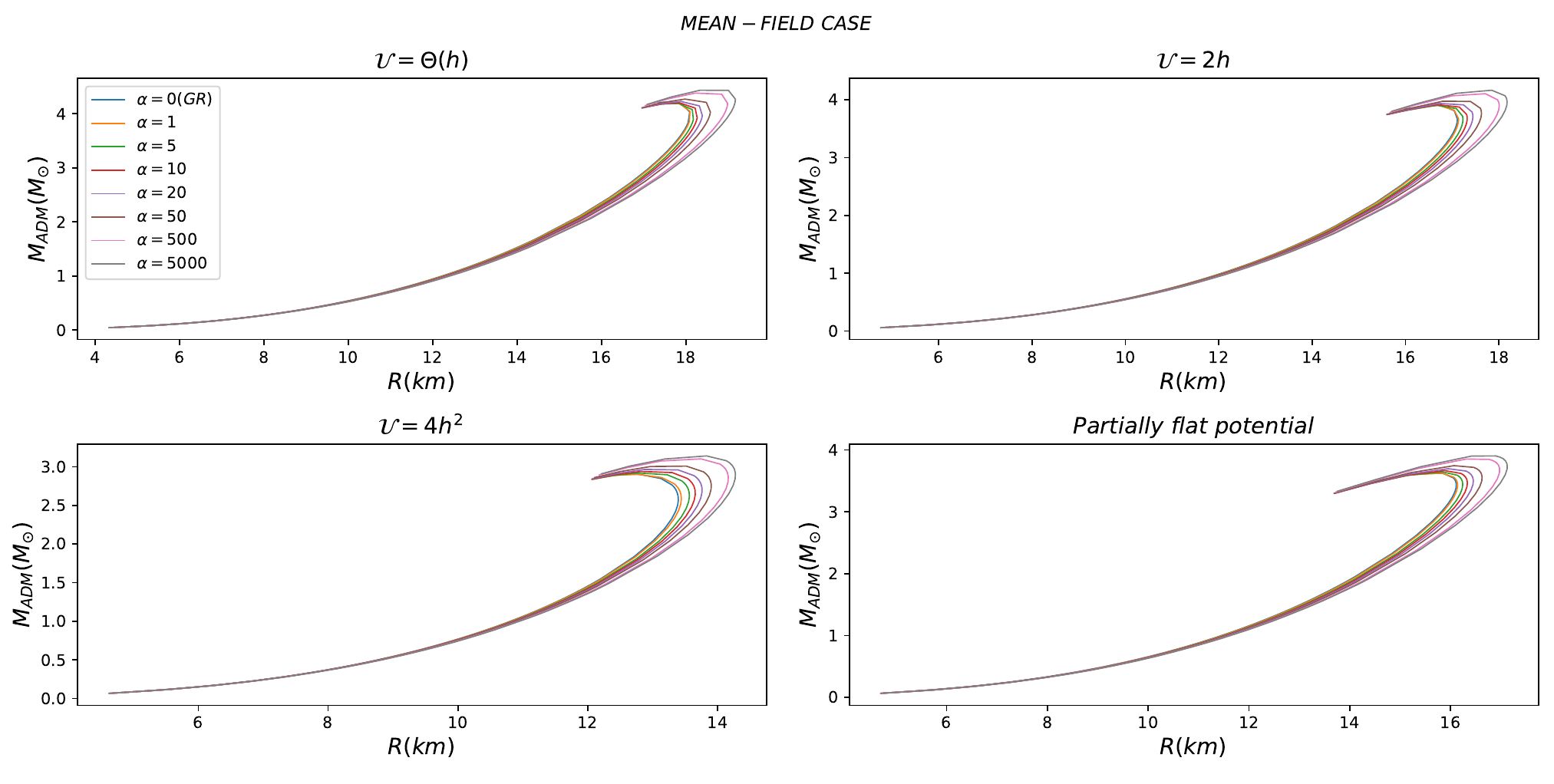}
	%\label{fig:MvsR_TOV}
	\caption{Mass-radius curve in the mean-field case, for the three potentials of Fig. 1 and for the constant potential, for several values of $\alpha$ (in units of km$^2$). 
For a given ADM mass, the star radius always grows with $\alpha$. The maximum mass grows slightly with increasing $\alpha$.} 
	\label{fig:MvsR_TOV}
\end{figure}

Another important issue in the study of neutron stars is the stability of the solution. The transition from stability to the unstable branch occurs at that value of the central density where the total energy (ADM mass) and the nucleon number are stationary \cite{Weinberg:1972kfs},
\begin{equation}
    \frac{dM}{d\rho_0}=0, \hspace{4mm} \frac{dN}{d\rho_0}=0,
\end{equation}
where $\rho_0$ is the central energy density. Then, we will find a maximum mass ($M_{max}$) beyond which the stars are no longer stable.

%% Figure 3
%\vspace{-1cm}
\begin{figure}%[H]
	%\centering
		\hspace{-0.0cm}\includegraphics[width=16cm, height=8cm]{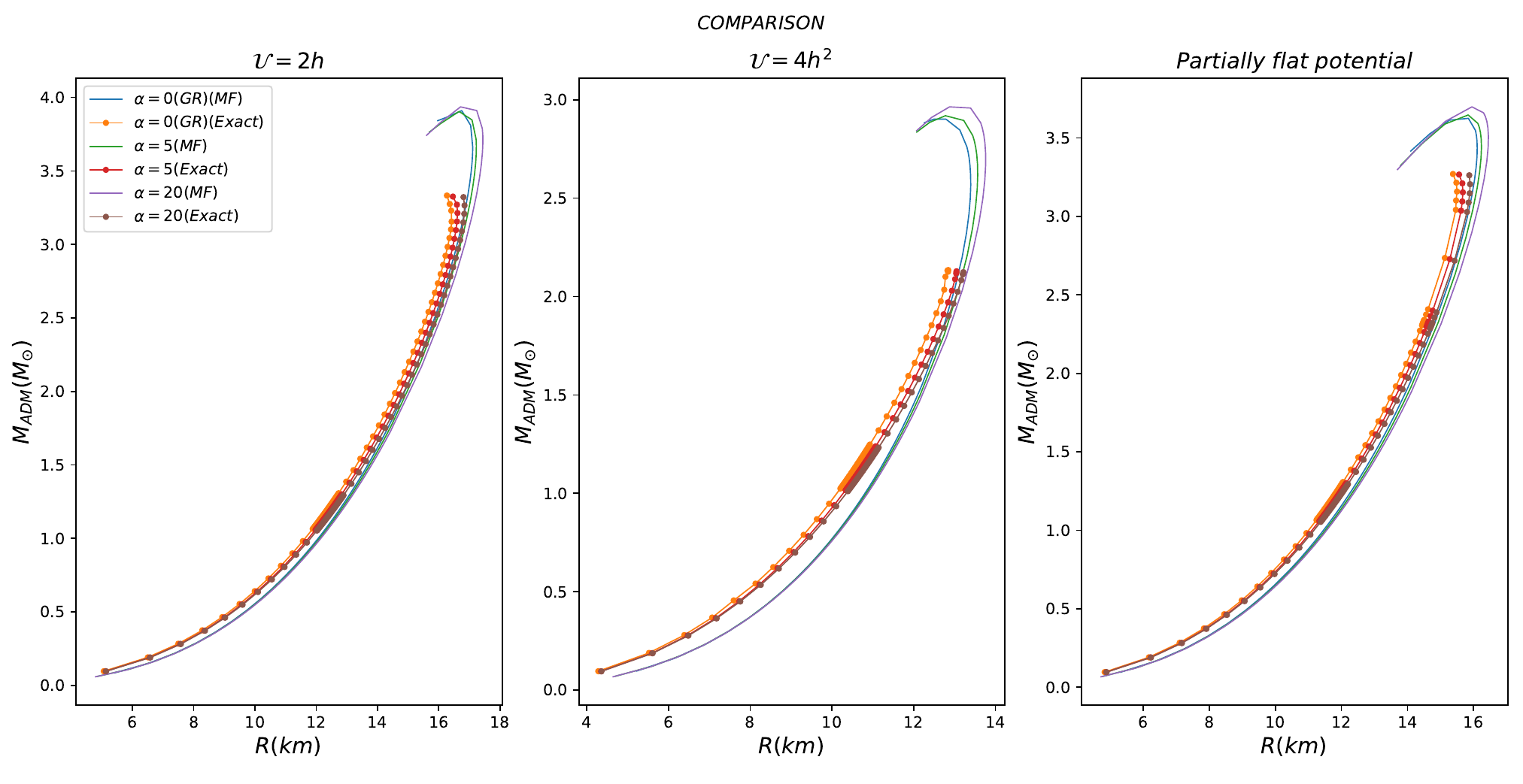}
	%\label{fig:MvsR_TOV}
	\caption{Comparison of the mass-radius curves between the exact and the mean-field case, for the three potentials of Fig. 1. The exact case always leads to a slightly more compact star and does not reach as high maximal masses as the mean-field case.}
	\label{fig:MvsR_comparison}
\end{figure}

We show the curves of the asymptotic (ADM) mass and the baryon number against the radius for both the exact and the mean-field case in Figs. 1-5. Further, in Figs. 6 and 7 we plot the mass against the central energy density.
Here, we measure the mass and baryon number in units of the solar mass $M_\odot$ and solar baryon number $N_\odot$, where
\begin{equation}
M_\odot = 1.998\cdot 10^{30}\, {\rm kg} \; , \quad N_\odot = \frac{M_\odot}{m_p} = 1.188 \cdot 10^{57} 
\end{equation}
and $m_p = 1.673 \cdot 10^{-27} \, {\rm kg} $ is the proton mass. Further, $\alpha$ is always given in units of km$^2 $.

We find that for stars with {\em small masses} (i.e., for sufficiently small central pressures or, equivalently, central energy densities), the radius and the value of the ADM mass for a fixed $p_0$ {\em decrease} with increasing $\alpha$ and, therefore, with respect to the GR case. The decrease in $M_{\rm ADM}$ can be directly seen in Figs. 6 and 7, whereas the corresponding decrease in the star radius follows from Figs. 1 and 2. For small masses, the $M(R)$ curves in Figs. 1 and 2 are almost identical for different $\alpha$, so smaller masses correspond to smaller radii. 
%% Figure 4
%\vspace{1cm}
\begin{figure}%[H]
	%\centering
		\hspace{-0.0cm}\includegraphics[width=16cm, height=10cm]{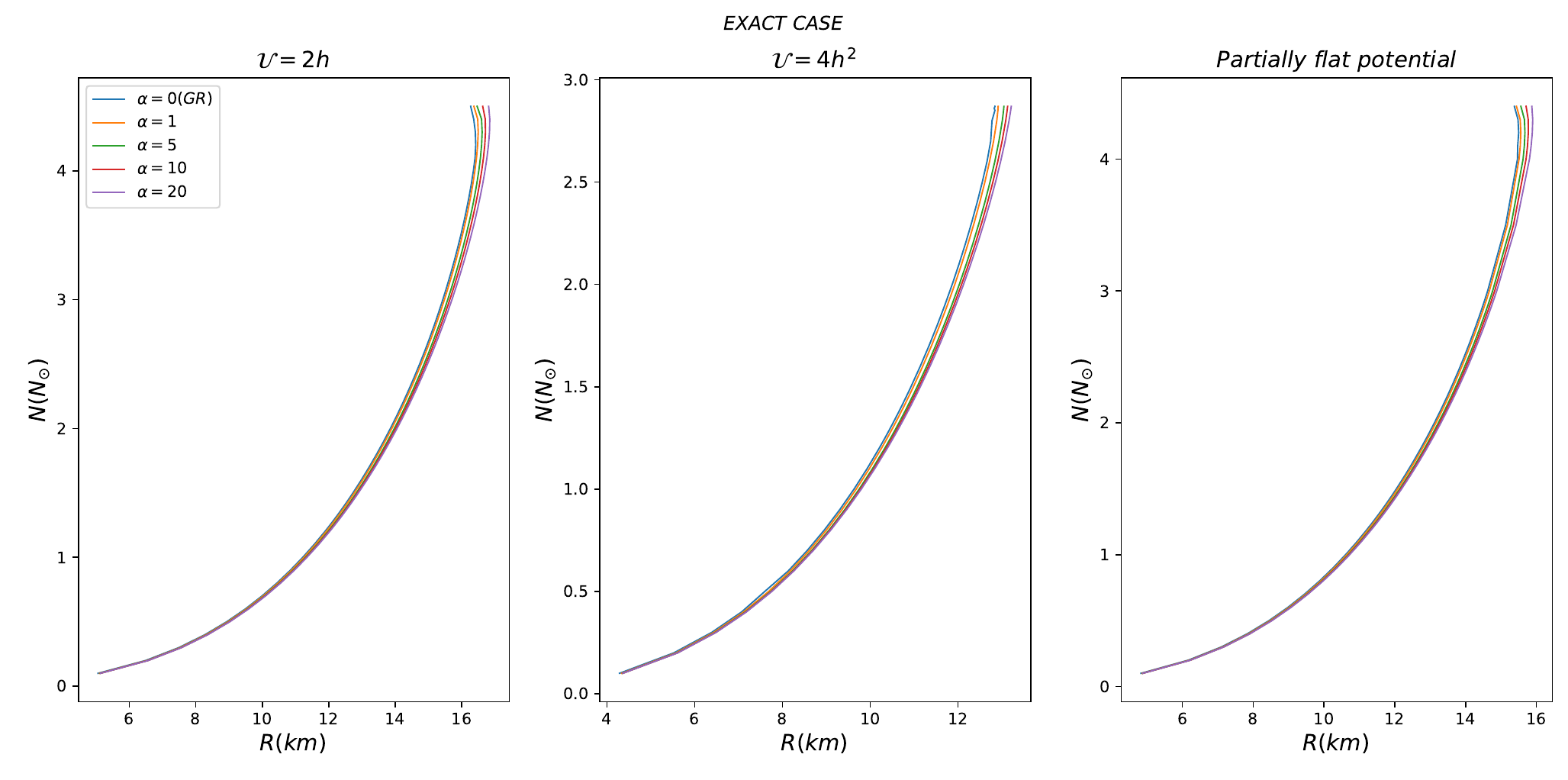}
		\caption{Baryon number $N$ vs. neutron star radius $R_s$, for the exact case.  It can be seen that the $N(R)$ curves are very similar to the $M(R)$ curves of Fig. 1.}
	\label{fig:NvsR_BPS}
\end{figure}
For higher values of the central pressure (corresponding to larger masses), the radii and the masses of the stars for fixed $p_0$ {\em increase} with increasing $\alpha$ (i.e., stronger deviations from GR). 
Again, the increase of the masses can be directly seen in Figs. 6 and 7, whereas the corresponding increase of the radii can be inferred from Figs. 1 and 2. For the maximum ADM masses we provide the corresponding values also in Tables 1 and 2.

\begin{table}[h!]
\begin{tabular}{|c|c|c|c|c|c|}\hline
\multicolumn{6}{|c|}{$2h$-Potential}\\ \hline
$\alpha$ $(\text{km}^2)$ & 0 & 1 & 5 & 10 & 20 \\ \hline
$M_{\text{ADM}}$ $(M_{\odot})$ & 3.332  & 3.327  & 3.325  &  3.323  &  3.321  \\ \hline
$R$ $(\text{km})$ & 16.26  & 16.36  & 16.46  & 16.63  & 16.81   \\ \hline
$\rho_0$ $(\text{MeV/fm}^3)$ & 820.73  & 817.64  & 768.10  & 704.61   &  655.20  \\ \hline
\end{tabular}
%\end{table}

\vspace*{0.2cm}

%\begin{table}[h!]
\begin{tabular}{|c|c|c|c|c|c|}\hline
\multicolumn{6}{|c|}{$4h^2$-Potential}\\ \hline
$\alpha$ $(\text{km}^2)$ & 0 & 1 & 5 & 10 & 20 \\ \hline
$M_{\text{ADM}}$ $(M_{\odot})$ & 2.134  & 2.130  & 2.128  &  2.126  &  2.124  \\ \hline
$R$ $(\text{km})$ & 12.84  & 12.91  & 13.05  & 13.13  & 13.21   \\ \hline
$\rho_0$ $(\text{MeV/fm}^3)$ & 2188.60  & 2116.41  & 1951.95  & 1872.46   &  1807.41  \\ \hline
\end{tabular}
%\end{table}

\vspace*{0.2cm}

%\begin{table}[h!]
\begin{tabular}{|c|c|c|c|c|c|}\hline
\multicolumn{6}{|c|}{Partially flat potential}\\ \hline
$\alpha$ $(\text{km}^2)$ & 0 & 1 & 5 & 10 & 20 \\ \hline
$M_{\text{ADM}}$ $(M_{\odot})$ & 3.270  & 3.268  & 3.266  &  3.264  &  3.262  \\ \hline
$R$ $(\text{km})$ & 15.37  & 15.44  & 15.56  & 15.71  & 15.87   \\ \hline
$\rho_0$ $(\text{MeV/fm}^3)$ & 603.44  & 605.00  & 557.36  & 516.89   &  482.39  \\ \hline
\end{tabular}
\caption{Exact case: values of the neutron star radii and the central energy densities for the maximum mass stars for the three potentials considered, for different values of $\alpha$. }
\end{table}

\begin{table}[h!]
\begin{tabular}{|c|c|c|c|c|c|c|c|}\hline
\multicolumn{8}{|c|}{$\Theta$-Potential}\\ \hline
$\alpha$ $(\text{km}^2)$ & 0 & 5 & 10 & 20 & 50 & 500 & 5000 \\ \hline
$M_{\text{ADM}}$ $(M_{\odot})$ & 4.218  & 4.219  &  4.221  &  4.227 &  4.275  &  4.402 & 4.447  \\ \hline
$R$ $(\text{km})$ & 17.58  & 17.60 & 17.79  & 17.92 & 18.08 & 18.55 & 18.95   \\ \hline
$\rho_0$ $(\text{MeV/fm}^3)$ & 430.70 & 374.21  & 360.09 &  345.97 & 338.91 & 317.73 & 289.49 \\ \hline
\end{tabular}
%\end{table}

\vspace*{0.2cm}

%\begin{table}[h!]
\begin{tabular}{|c|c|c|c|c|c|c|c|}\hline
\multicolumn{8}{|c|}{$2h$-Potential}\\ \hline
$\alpha$ $(\text{km}^2)$ & 0 & 5 & 10 & 20 & 50 & 500 & 5000 \\ \hline
$M_{\text{ADM}}$ $(M_{\odot})$ & 3.918  & 3.919  &  3.921  &  3.944 & 3.990  &  4.113 & 4.170  \\ \hline
$R$ $(\text{km})$ & 16.56  & 16.70  & 16.82 & 16.93 & 17.09 & 17.50 & 17.69   \\ \hline
$\rho_0$ $(\text{MeV/fm}^3)$ & 489.46 & 426.80  & 408.83  &  399.83 & 390.82 & 372.75 & 363.70 \\ \hline
\end{tabular}
%\end{table}

\vspace*{0.2cm}

%\begin{table}[h!]
\begin{tabular}{|c|c|c|c|c|c|c|c|}\hline
\multicolumn{8}{|c|}{$4h^2$-Potential}\\ \hline
$\alpha$ $(\text{km}^2)$ & 0 & 5 & 10 & 20 & 50 & 500 & 5000 \\ \hline
$M_{\text{ADM}}$ $(M_{\odot})$ & 2.905  & 2.921  &  2.944  &  2.974 & 3.019  &  3.106 & 3.143  \\ \hline
$R$ $(\text{km})$ & 12.63  & 12.90  & 13.00 & 13.12 & 13.29 & 13.61 & 13.69   \\ \hline
$\rho_0$ $(\text{MeV/fm}^3)$ & 825.34 & 705.17 & 689.90 & 674.56 & 659.15 & 628.06 & 596.59  \\ \hline
\end{tabular}
%\end{table}

\vspace*{0.2cm}

%\begin{table}[h!]
\begin{tabular}{|c|c|c|c|c|c|c|c|}\hline
\multicolumn{8}{|c|}{Partially flat potential}\\ \hline
$\alpha$ $(\text{km}^2)$ & 0 & 5 & 10 & 20 & 50 & 500 & 5000 \\ \hline
$M_{\text{ADM}}$ $(M_{\odot})$ & 3.63  &  3.64 & 3.66 &  3.69 & 3.74 &  3.85 & 3.90  \\ \hline
$R$ $(\text{km})$ & 15.627 & 15.77 & 15.88 & 16.07 & 16.31 & 16.72 & 16.89   \\ \hline
$\rho_0$ $(\text{MeV/fm}^3)$ & 484.492 & 447.23 & 434.77 & 414.78 & 400.98 & 378.337 & 372.02  \\ \hline
\end{tabular}
\caption{Mean-field case: values of the neutron star radii and the central energy densities for the maximum mass stars for the four potentials considered, for different values of $\alpha$.}
\end{table}

From Figs. 1, 2 it appears as if the $M(R)$ curves for different $\alpha$ approached each other for small masses. This is, however, not entirely correct. The different $M(R)$ curves for different $\alpha$ for a given model, in fact, always cross each other in the region of small $M$. In particular, for each $\alpha$ there exists a neutron star mass $M_*(\alpha)$ which has exactly the same radius as its GR counterpart ($\alpha =0$). For all our models, however, this occurs for very small masses (always smaller than $0.15 \, M_\odot$). As such small masses are most likely phenomenologically irrelevant, we did not try to zoom into this region to make this behavior more visible. The fact that this crossing of different $M(R)$ curves happens for very small masses is probably related to the very stiff nature of our EoS even for small density. 
BPS Skyrme neutron stars, by construction, do not have a crust region, although a crust can be added without difficulty \cite{Adam2020crust}.  
Other EoS, which are much softer in the low-density region, produce pronounced crust regions for small mass neutron stars and, thus, the crossing happens for much larger masses, see, e.g., \cite{Sbisa:2019mae}.     
%% Figure 5
%\vspace{-1cm}
\begin{figure}%[H]
	%\centering
		\hspace{-0.0cm}\includegraphics[width=16cm, height=10cm]{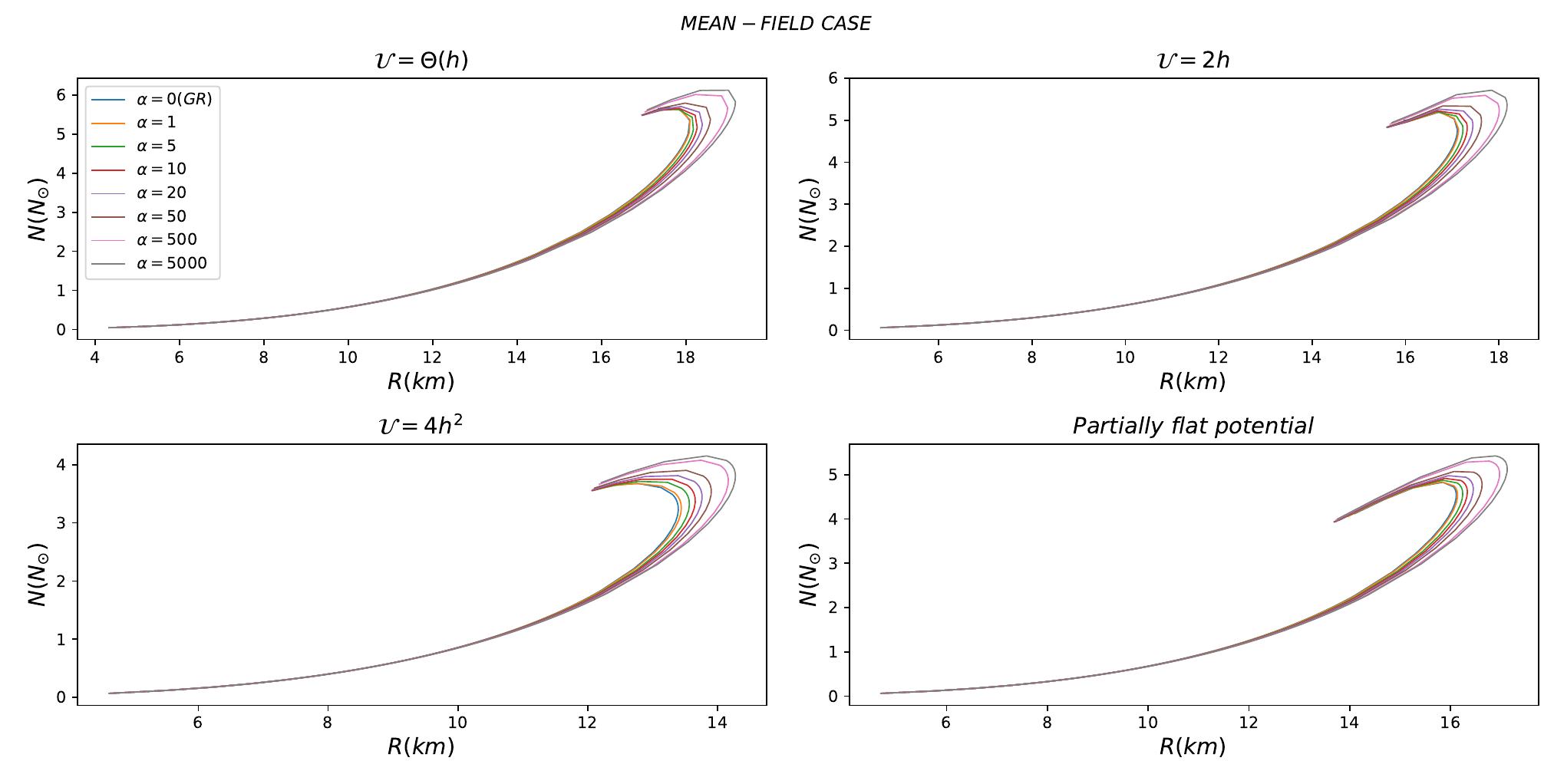}
		\caption{Baryon number $N$ vs. neutron star radius $R_s$, for the mean-field case. }
	\label{fig:NvsR_TOV}
\end{figure}

Another interesting quantity is the mass at the surface of the star, $M_s$, which in $f(R)$ gravity is a second, independent and invariant mass observable, as explained in \cite{Sbisa:2019mae}. It may be understood as a sum of the mass contributions of matter and curvature inside the star.  As in \cite{astashenok2017realistic}, we find that its value decreases when we deviate from GR (i.e., for increasing $\alpha$), see Figs. 8, 9. It turns out, however, that the non-vanishing curvature scalar outside the star produces a further contribution to the ADM mass which essentially compensates this decrease.  
The region outside the star, in which the Ricci scalar does not vanish, is also referred to as the \textit{gravisphere} \cite{astashenok2017realistic}, and it can be seen explicitly in Fig. 10. We only include the mean-fields plots because they are very similar to the exact case.

An even further mass definition (the "Newtonian mass" $M_n$ \cite{Sbisa:2019mae}) is provided by the time-time metric function $A(r)$,
\begin{equation}
M_n(r) = \frac{4\pi r}{\kappa} \left( 1 - A (r) \right) \; , \quad M_{n,s} \equiv M_n (R_s) ,
\end{equation}
and this mass is relevant for the surface redshift $z_s$ \cite{Sbisa:2019mae}, 
\begin{equation}
	z_s = \left( 1-\frac{\kappa M_{n,s}}{4\pi R_s} \right)^{-1/2} -1.
\end{equation}
Interestingly, it turns out that $M_{n,s}$ is {\em larger} than $M_{\rm ADM}$ such that the value of $z_s$ for a star of a given mass {\em increases} in comparison to the one predicted in GR, see Figs. 11-13. This is probably related to the particularly stiff nature of the EoS of the BPS Skyrme model, because for the soft EoS used in \cite{Sbisa:2019mae} they find the opposite behavior. In any case, this difference could be important to discriminate between different extended theories of gravity. 
%% Figure 6
%\vspace{1cm}
\begin{figure}%[H]
	%\centering
		\hspace{-0.0cm}\includegraphics[width=16cm, height=10cm]{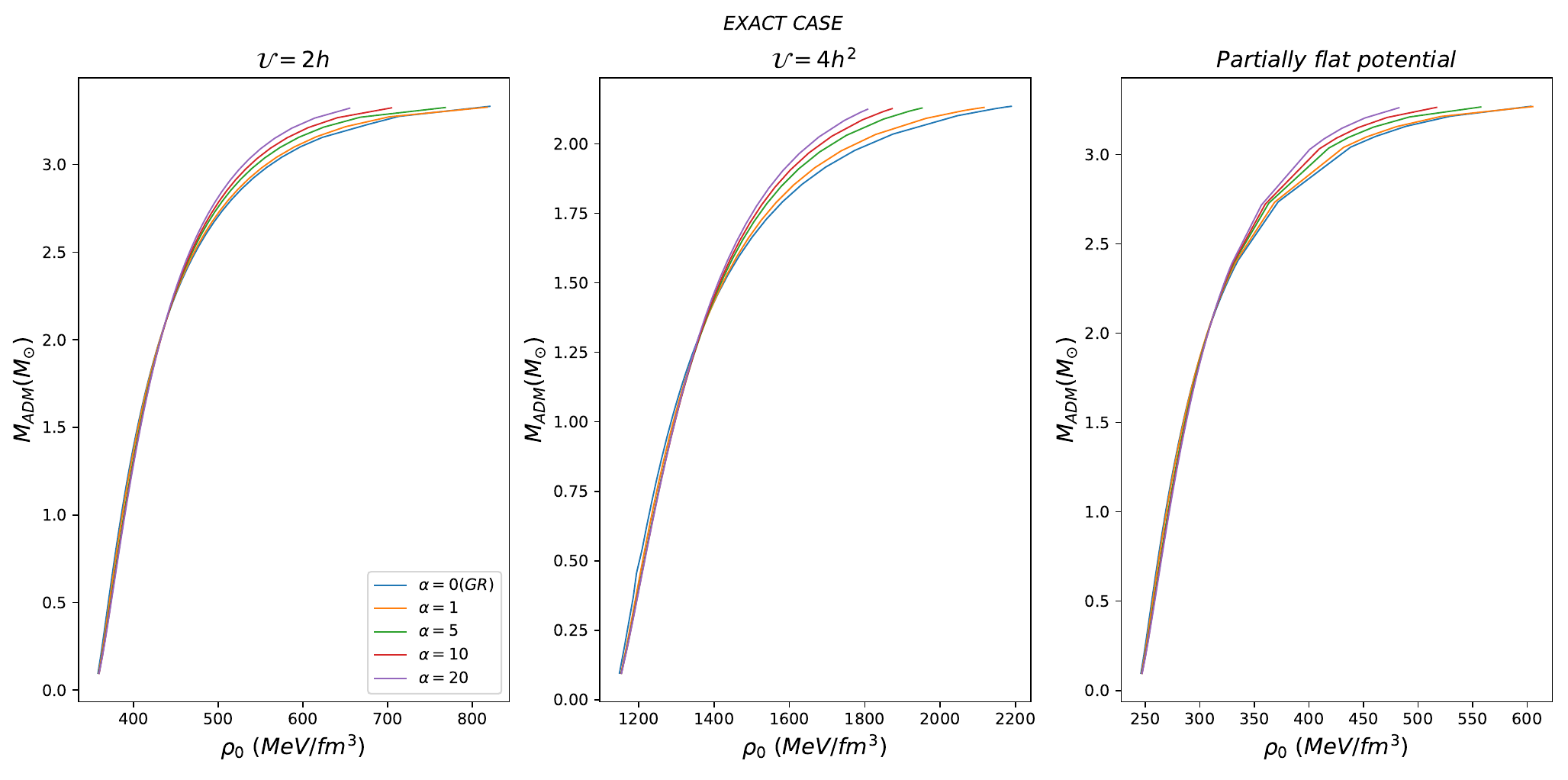}
		\caption{The ADM mass as a function of the central energy density $\rho_0$, for the exact case. It can be clearly seen that for a fixed $\rho_0$ the mass decreases with increasing $\alpha$ for small masses, but increases with increasing $\alpha$ for large masses.  }
	\label{fig:Mvsrho_BPS}
\end{figure}

The Ricci scalar curves for the mean-field case shown in Fig. 10 are continuous at the neutron star radius for the modified gravity $\alpha >0$, but discontinuous for the GR case. This discontinuity is a consequence of the EoS of the BPS model in the mean-field case because, in GR, the Ricci scalar obeys a purely algebraic (constraint) equation given by the trace of the Einstein equations,
\begin{equation} \label{alg-ricci}
	R = \kappa (3p - \rho).
\end{equation}
Therefore, if the EoS leads to a non-vanishing energy density at the neutron star's surface, the Ricci scalar will also show that discontinuity. In the modified gravity, on the other hand, the curvature satisfies its own differential equation until the end of the integration, which results in a continuous curve.

Finally, we want to comment on a possible singularity which, interestingly, is always avoided by solutions of our system. Indeed, the first derivative of $f(R)$,
\begin{equation}
f_R (R) = 1-2\alpha R,
\end{equation}
may, in principle, become zero for a positive $R$, which would introduce a singularity in the system of equations (\ref{eqB}) - (\ref{eqR}). In GR, $R$ is always negative close to the surface, where $\rho$ dominates over $p$, see Eq. (\ref{alg-ricci}). Whether it may become positive in the center of the star depends on the EoS. It may become positive in our case, because in the high-pressure limit the EoS of the BPS Skyrme model approximates the maximally stiff EoS $\rho = p \, +\,$const. Indeed, it can be seen in Fig. 8 that $R(0)$ takes positive values in the GR case ($\alpha =0$). On the other hand, we also see that $R(0)$ {\em diminishes} with increasing $\alpha$. In particular, it seems that $R(0)$ always becomes negative for sufficiently large $\alpha$. In any case, $f_R (R(r)) = 1-2\alpha R(r)$ always remains positive for all $r$ for all solutions we considered (even for extremely large $\alpha$), and the singularity never occurs.
%% Figure 7
%\vspace{-1cm}
\begin{figure}%[H]
	%\centering
		\hspace{-0.0cm}\includegraphics[width=16cm, height=10cm]{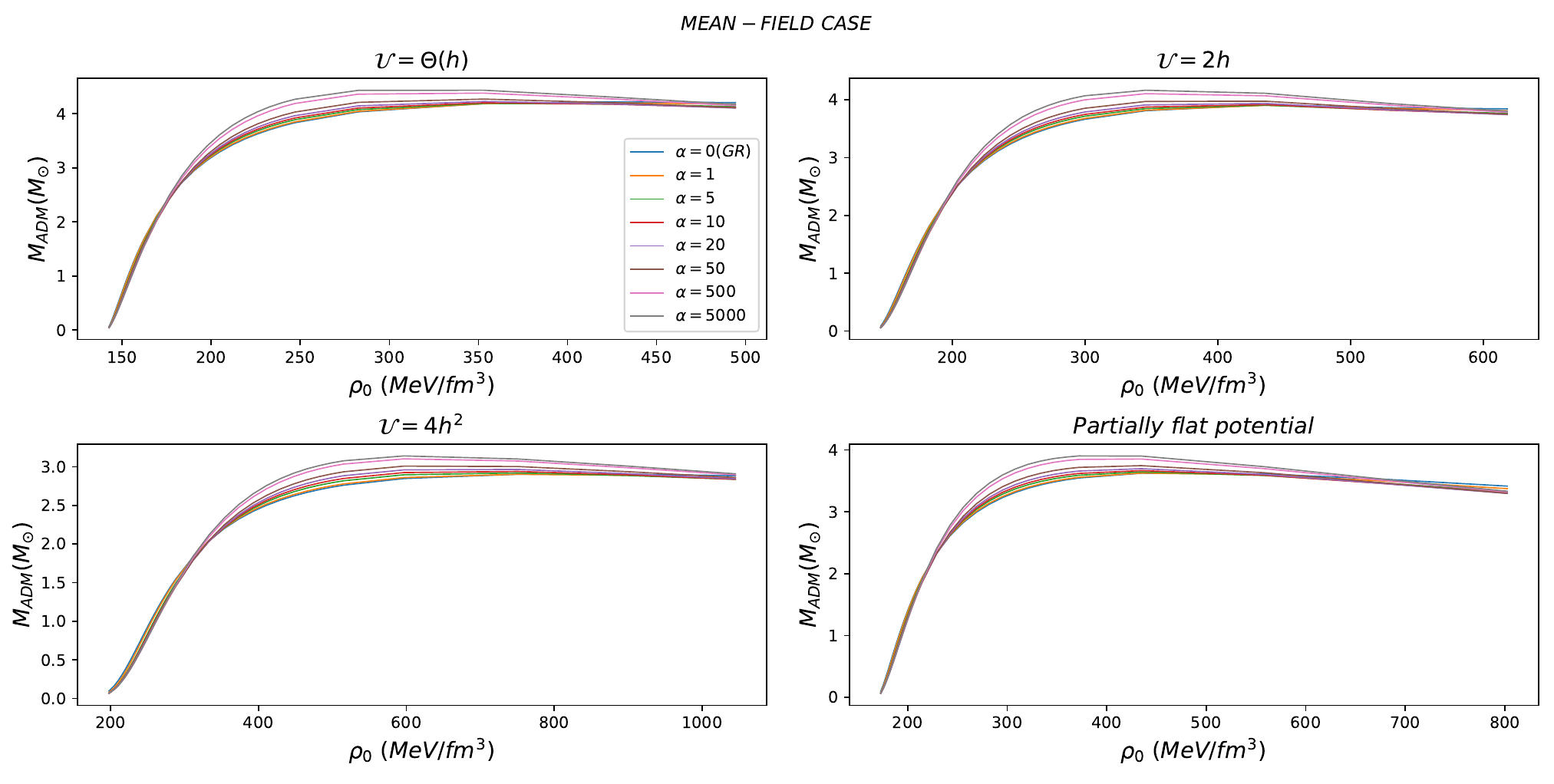}
		\caption{The ADM mass as a function of the central energy density $\rho_0$, for the mean-field case. It can be clearly seen that for a fixed $\rho_0$ the mass decreases with increasing $\alpha$ for small masses, but increases with increasing $\alpha$ for large masses. The endpoints of the curves correspond to the value of $p_0$ where the GR case ($\alpha =0$) reaches its maximum mass. This implies that the endpoints of the other curves are beyond their maxima and already belong to the unstable branch.}
	\label{fig:Mvsrho_TOV}
\end{figure}

\section{Conclusions}
The main aim of this paper was to study the principal differences that result when coupling the BPS Skyrme model to a $f(R)$ theory. Concretely, we chose the Starobinsky model as the simplest possibility. Similar studies have already been performed with other EoS, where always one shooting method is involved. The resolution of the BPS model in the exact case, on the other hand, does imply a shooting method even in GR, so a double shooting method is required in the $f(R)$ case. In general, our results are compatible with those obtained in other investigations \cite{astashenok2017realistic,kase2019neutron,Sbisa:2019mae,deliduman2012neutron,staykov2014slowly,Astashenok:2014dja}.
In more detail, our $M(R)$ curves are quite similar to the curves resulting from quark stars \cite{Astashenok:2014dja}, although the deviations for different values of $\alpha$ are slightly larger in the quark star case (probably related to the fact that the quark star EoS for high densities is softer than the BPS Skyrme EoS). The underlying reason for their similarity is that in both cases (quark stars and BPS Skyrme stars) the EoS does not become extremely soft in the low-density region, such that the $M(R)$ curves always have positive slope for small masses. In other words, the radius grows with the mass, and light stars do not have a pronounced tail (or crust). 
For EoS which approach the very soft EoS of nuclear physics for low densities, on the other hand, the resulting $M(R)$ curves lead to larger radii for smaller masses (a negative slope) in the low-density region. For such EoS, the effect of varying $\alpha$ is much stronger, particularly for small mass neutron stars \cite{astashenok2017realistic,Sbisa:2019mae}. There seems to exist an overall tendency that the variation of $\alpha$ has a stronger effect for softer EoS.
%% Figure 8
\begin{figure}%[H]
	%\centering
		\hspace{-0.0cm}\includegraphics[width=16cm, height=10cm]{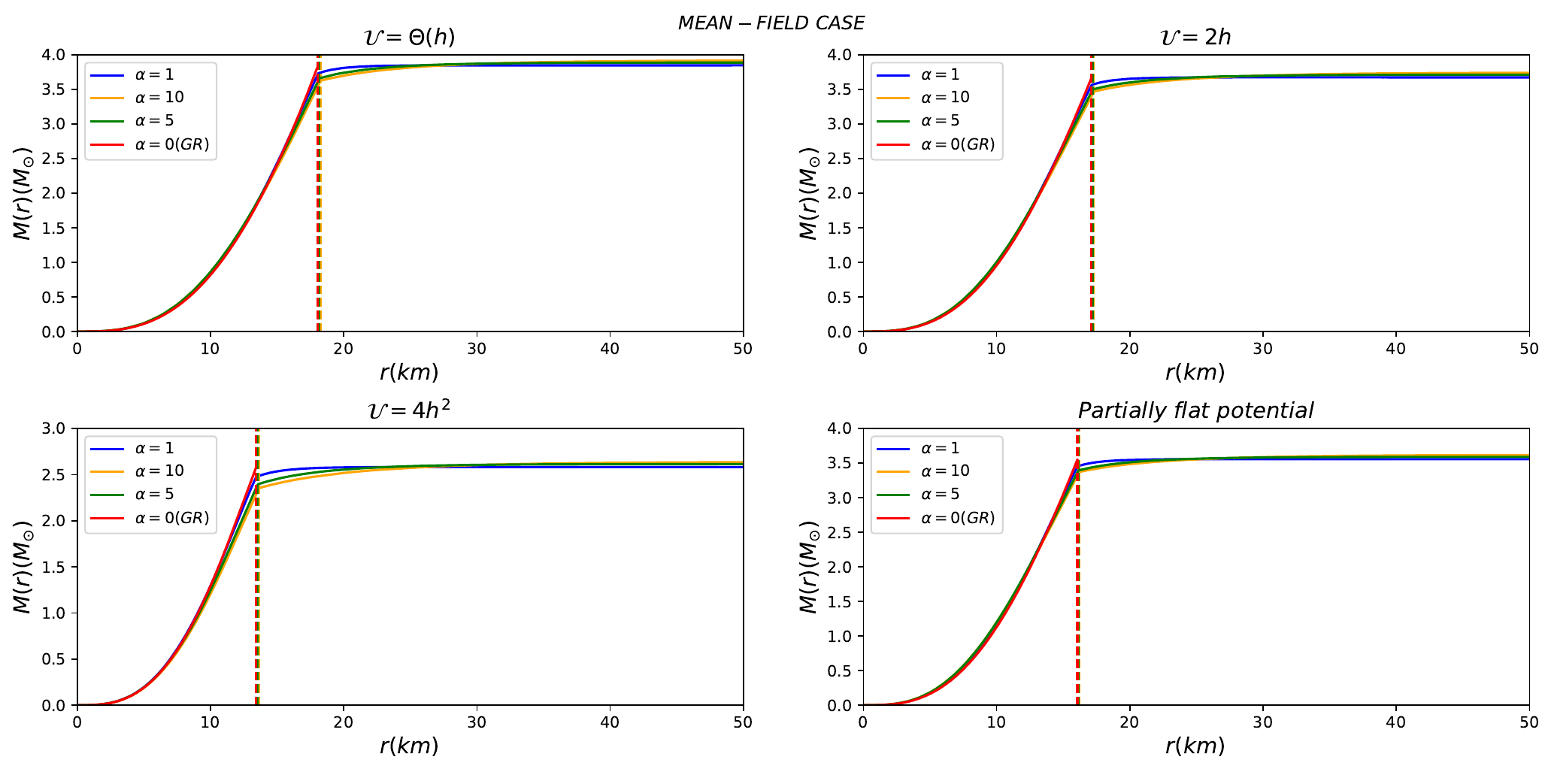}
		\caption{Mass function $M(r)$ vs radius, for the mean-field case. For each potential, the solutions are for a fixed value of $p_0$ which is sufficiently large to lead to a rather large mass but, at the same time, sufficiently small such that all solutions belong to the stable branch, even for $\alpha = 10$. It can be seen that the mass at the neutron star radius, $M_s = M(R_s)$, decreases with increasing $\alpha$. On the other hand, the mass remains constant for $r>R_s$ in the GR case but continues to grow for $\alpha >0$, such that the asymptotic or ADM mass is slightly larger for larger values of $\alpha$. In the GR case, we do not show the constant curve for $r>R_s$, because the integration stops there. The vertical dashed lines indicate the neutron star radii for different $\alpha$, which turn out to be very similar.}
	\label{fig:massvsrTOV}
\end{figure}

%% Figure 9
%\vspace{-1cm}
\begin{figure}%[H]
	%\centering
		\hspace{-0.0cm}\includegraphics[width=16cm, height=10cm]{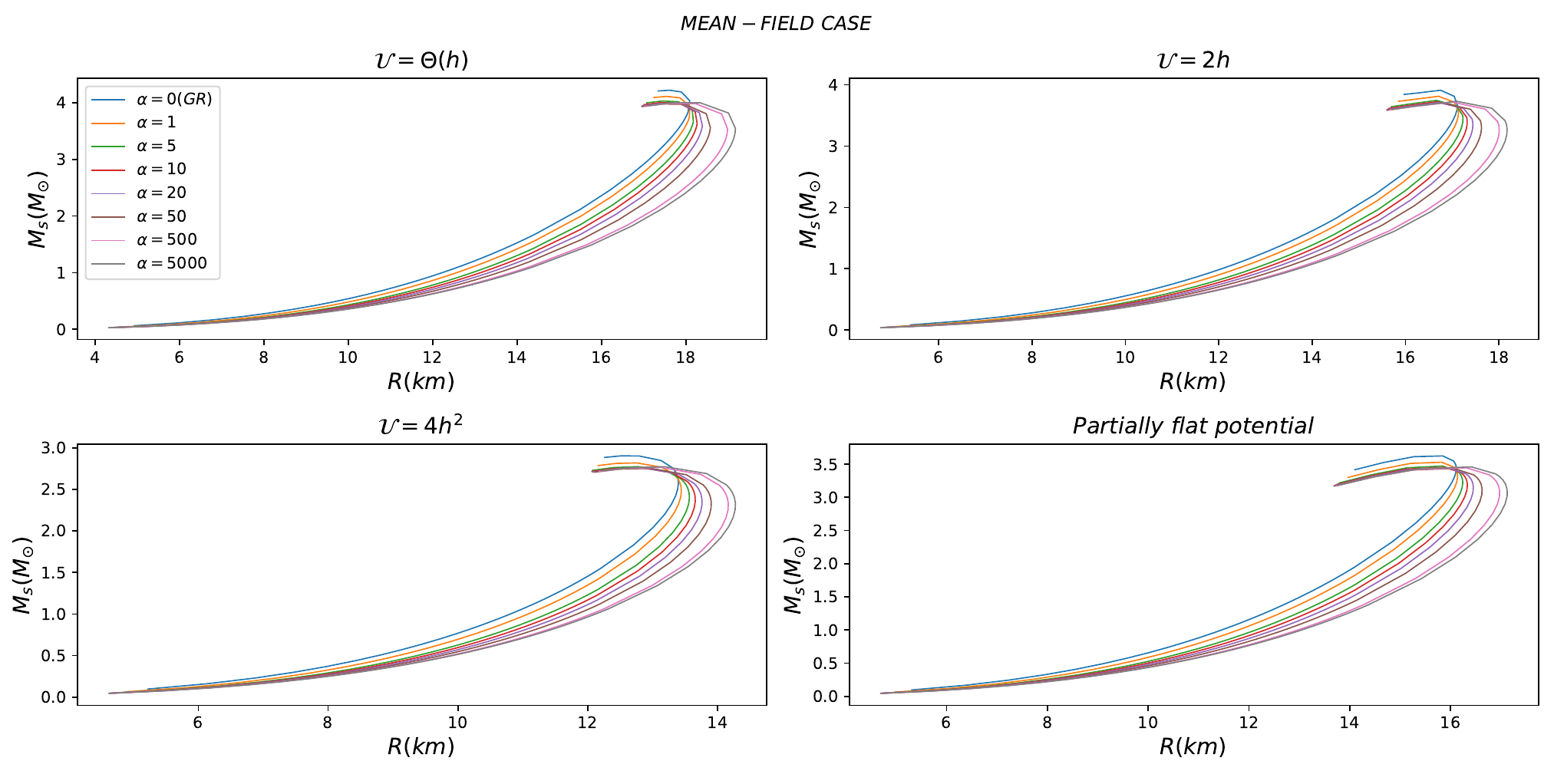}
		\caption{The surface mass $M_s$ as a function of the star radius $R_s$, for the mean-field case. It can be clearly seen that the surface mass for a given radius shrinks with increasing $\alpha$. }
	\label{fig:MsvsR_TOV}
\end{figure}

It is interesting to compare the values of $\alpha$ (the parameter of the Starobinsky model) used in this paper (which are similar to the values used, e.g., in \cite{astashenok2017realistic,Sbisa:2019mae,Astashenok:2014dja}) to some observational astrophysical bounds. In  \cite{arapouglu2011constraints} a bound $\alpha \lesssim 1\, {\rm km}$ was suggested, based on bounds on neutron star masses. That bound, however, was derived within a perturbative approach to the Starobinsky model which is not capable of reproducing the gravisphere contribution to the neutron star mass and, therefore, underestimates this mass. Taking into account this correction, the bound becomes much weaker \cite{Astashenok:2014dja}. Other astrophysical bounds \cite{Naf:2010zy} are much weaker, as well, such that the values considered in the present article are, at this moment, compatible with those astrophysical bounds.
Further, we restricted to positive values of $\alpha$ to avoid tachyonic instabilities. As explained in section II, in the case of $\alpha<0$ we find solutions of the Ricci scalar that show damped oscillations  outside the star. A more detailed investigation of this case can be found in \cite{resco2016neutron} and some comments in \cite{astashenok2017realistic}.

For the neutron star observables we find that, for massive stars, the radius increases with $\alpha$, while the maximum mass also slightly increases in the mean-field case, see Table 2. In the exact case, the maximum ADM mass seems to slightly decrease with increasing $\alpha$, although the effect is tiny (see Table 1). So, in principle, we could constrain the values of $\alpha$ with observational neutron star data. However, such data are still not very precise, owing to the smallness and the lack of electromagnetic radiation emission of these objects. Besides, the maximum masses that the neutron stars can reach strongly depend on the EoS that we are using, hence those EoS that do not reach the minimum value required by experimental data ($\sim 2 M_{\odot}$ \cite{demorest2010shapiro}), but only fail to do so by a small amount, could be reconsidered with these results. A priori one might think that the maximum mass can be arbitrarily large when the value of $\alpha$ is increased, but we find that this is not the case (e.g., for a value of $\alpha = 5000 \: \text{km}^2$ we find an increase of about 5 \%). This is also found in \cite{staykov2014slowly}.

%% Figure 10
\begin{figure}%[H]
	%\centering
		\hspace{-0.0cm}\includegraphics[width=16cm, height=10cm]{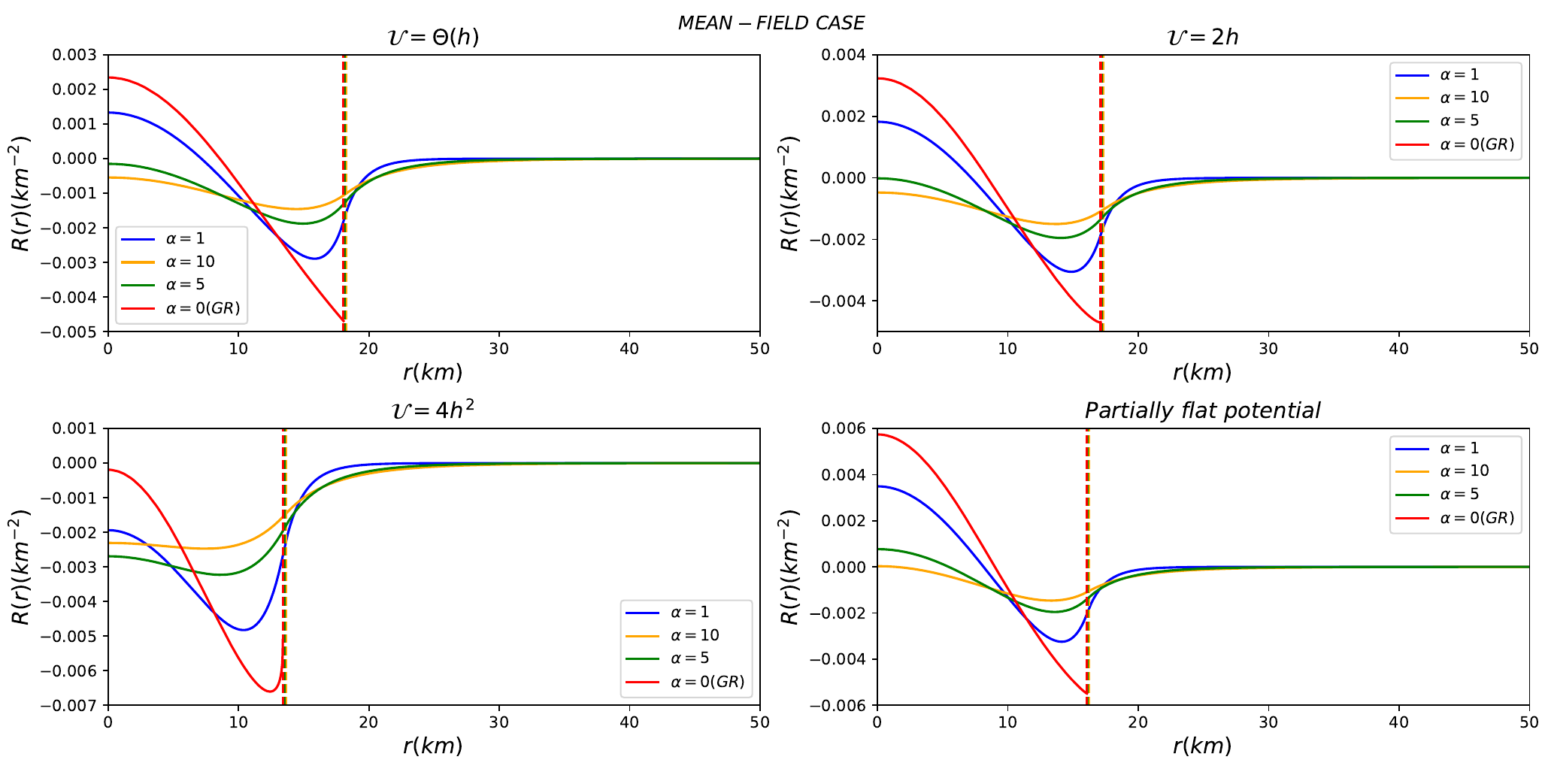}
		\caption{The Ricci scalar $R$ as a function of the radius, for the mean-field case. The solutions are for the same fixed values of $p_0$ as in Fig. 7. The Ricci scalar suddenly jumps from a (negative) non-zero value to zero in the GR case,  but is continuous at $R_s$ for $\alpha >0$. Further, it can be observed that, whenever $R(0)$ is positive, it decreases with increasing $\alpha$. In the GR case, we do not show the constant curve for $r>R_s$, because the integration stops there. The vertical dashed lines indicate the neutron star radii for different $\alpha$, which turn out to be very similar.}
	\label{fig:RicvsrTOV}
\end{figure}
One further interesting result is that the Newtonian surface mass $M_{n,s}$ relevant for the redshift of radiation emitted from the star surface is {\em larger} than the ADM mass, in contrast to results for softer EoS. This implies that if generalized gravity turns out to be indeed relevant for neutron stars, then the redshift will be able to distinguish different EoS and, in particular, their stiffness.

We also find that the Ricci scalar is smooth at the surface of the stars. In the GR case, an EoS leading to a non-vanishing energy density at the surface of the star leads to a discontinuity in the Ricci scalar, but in $f(R)$ gravity that discontinuity is cured, because $R$ satisfies its own differential equation. We required that the Ricci scalar tends to 0 at infinity to recover an asymptotically flat spacetime but, in principle, this would not be necessary, and we could have imposed the Schwarzschild solutions just at the surface of the star. This has been done in \cite{ganguly2014neutron,nzioki2014jebsen} for some EoS, but they find that this matching condition cannot be imposed for an arbitrary EoS, hence this matching is highly unnatural. The results shown in this paper are, thus, a straightforward violation of the Jebsen-Birkhoff theorem in $f(R)$ gravity.

%% Figure 11
%\vspace{-1cm}
\begin{figure}%[H]
	%\centering
		\hspace{-0.0cm}\includegraphics[width=16cm, height=10cm]{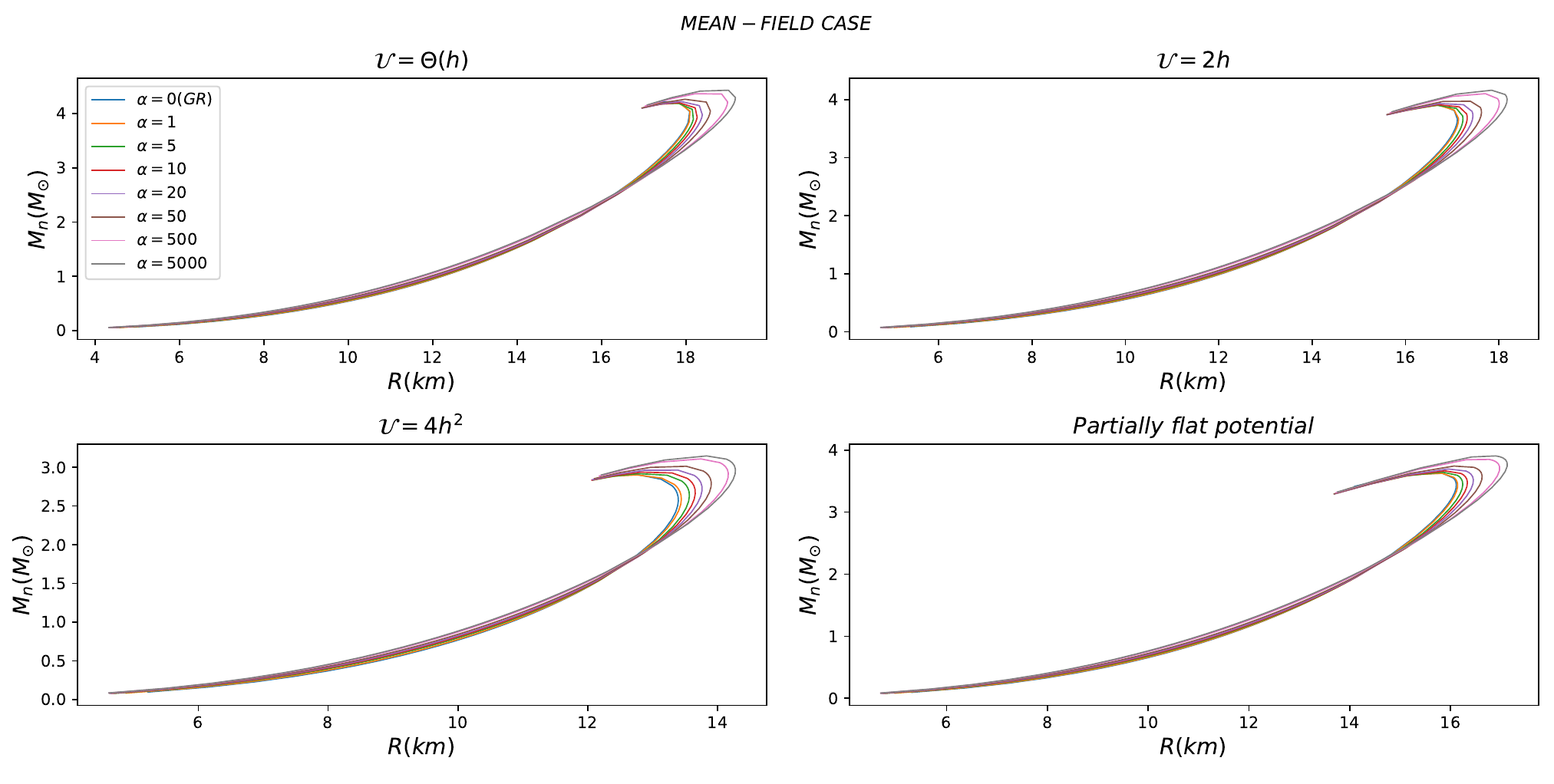}
		\caption{The Newtonian mass at the surface, $M_{n,s}$ as a function of the star radius, for the mean-field case. For a fixed radius, it grows with $\alpha$ for small stars but shrinks for large stars.}
	\label{fig:MnvsR_TOV}
\end{figure}

%% Figure 12
%\vspace{-1cm}
\begin{figure}%[H]
	%\centering
		\hspace{-0.0cm}\includegraphics[width=16cm, height=10cm]{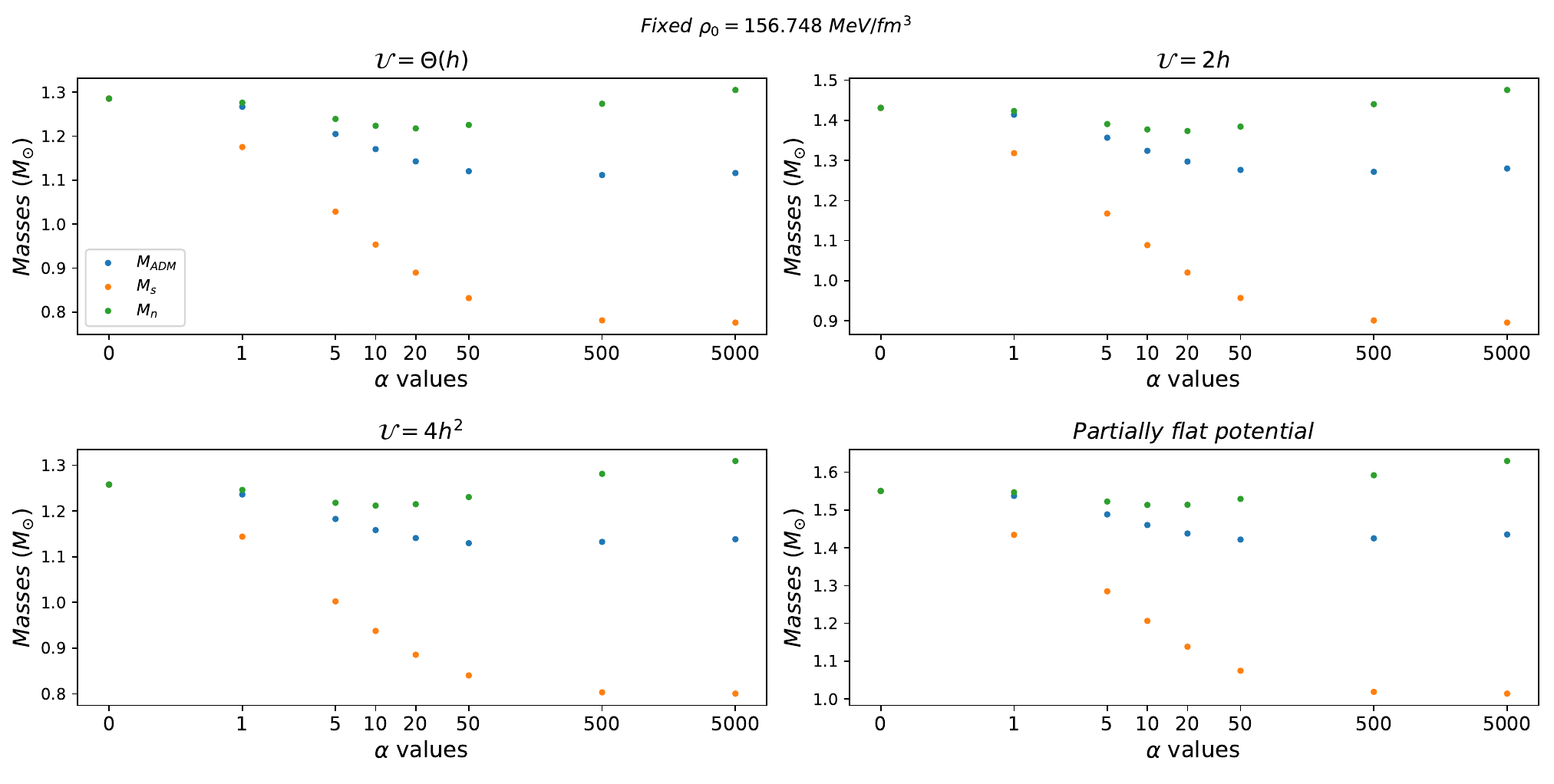}
		\caption{Comparison of the different masses, for different values of $\alpha$ and a fixed central density, in the mean-field case. It can be clearly seen that the Newtonian mass at the surface is larger than the ADM mass.}
	\label{fig:MassesvsAlpha_TOV}
\end{figure}

%% Figure 13
%\vspace{-1cm}
\begin{figure}%[H]
	%\centering
		\hspace{-0.0cm}\includegraphics[width=16cm, height=10cm]{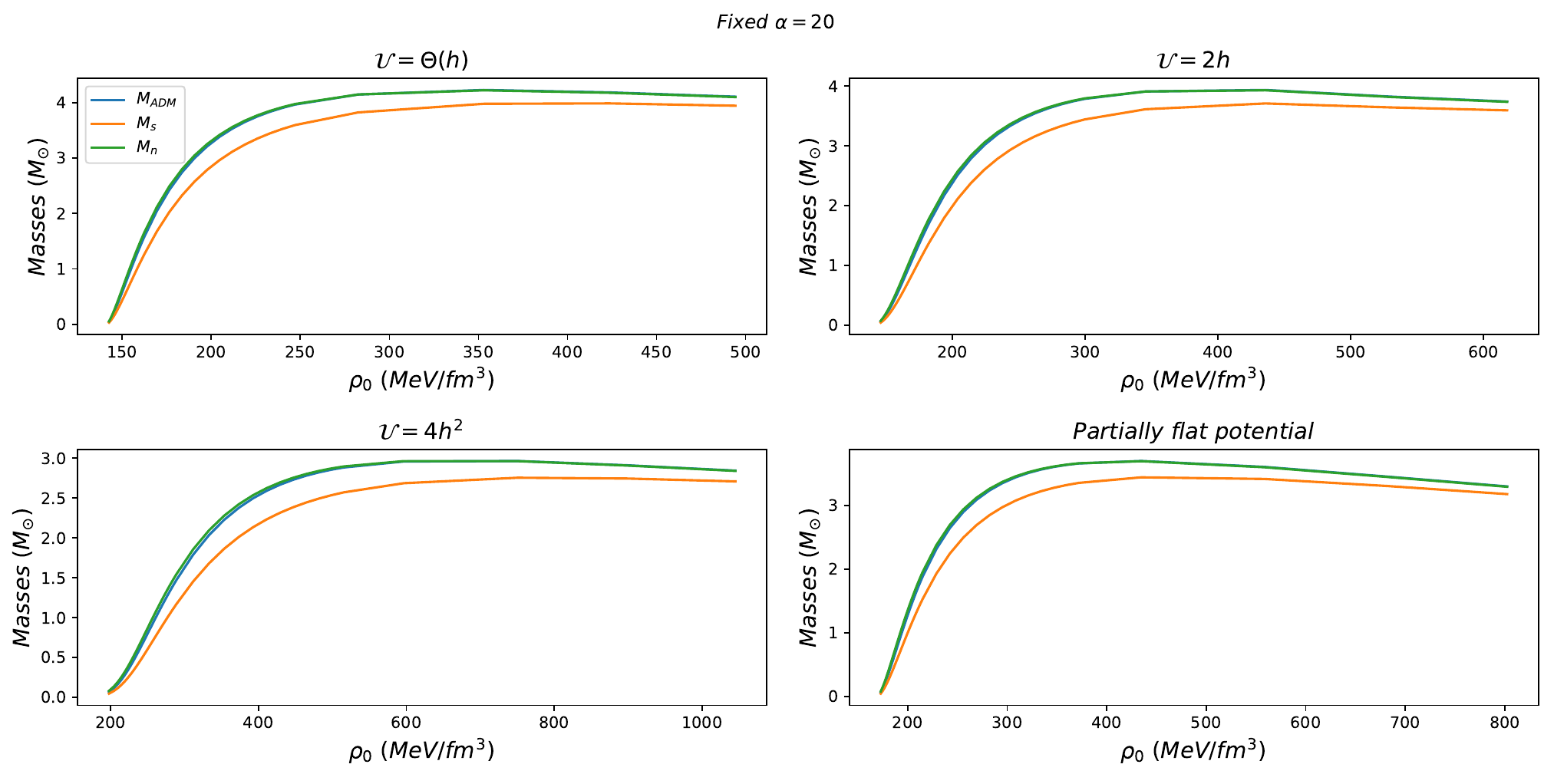}
		\caption{Comparison of the different masses, for different values of the central energy density $\rho_0$ and a fixed $\alpha$, in the mean-field case. It can be clearly seen that the Newtonian mass at the surface is larger than the ADM mass, particularly for an intermediate range of $\rho_0$.}
	\label{fig:MassesvsRho_TOV}
\end{figure}

These results also constitute an additional motivation to keep studying the BPS Skyrme model to describe neutron stars. A possible next step consists in the study of the differences of these results performed in the Einstein frame as is done in \cite{kase2019neutron}, although they present their final results in the Jordan frame. The Jordan frame, which is the one used in this paper, is the one in which General Relativity is usually expressed, but the introduction of modified theories of gravity, concretely the scalar-tensor theories, motivates the introduction of the Einstein frame via a conformal transformation of the metric.

Further investigations will be focused on rotating neutron stars, for which a set of relations (called the I-Love-Q relations \cite{yagi2013love}) has been found to be independent of the EoS. It is an interesting question to check the universality of these relations also with the BPS Skyrme model EoS. Further, universal relations like the I-Love-Q relation could help to constrain alternative theories of gravity. 
The EoS of dense nuclear matter is not completely determined, and the observational data of neutron stars are still not sufficient to pin down this EoS. We are, therefore, adding even more uncertainties when considering modified theories of gravity, and an EoS-independent relation could help to discard some of these modified theories.

In any case,
the BPS Skyrme model fitted to infinite nuclear matter does yield rather high values for the maximum mass even in GR (up to $\sim 3.5 M_{\odot}$ for realistic potentials), and generalized gravity does not help in this respect. For a more complete and more reliable description of nuclear matter, therefore, the BPS Skyrme submodel should be combined with the standard Skyrme model, which is known to lead to a smaller value of the maximum mass \cite{nelmes2012phase,Naya:2019rlm}. The investigation of the full generalized Skyrme model and its resulting neutron stars in $f(R)$ gravity is, therefore, an important next step in this investigation. It is expected that the BPS Skyrme submodel still will provide the leading contribution in the central region of the stars, because the sextic term $\mathcal{L}_6$ is known to essentially determine the EoS in the limit of large density \cite{Adam:2015lra}. The resulting model will then permit even more direct comparisons with observational constraints on the $M(R)$ curves (and other observables), like the ones obtained from photospheric radius expansions \cite{Steiner:2012xt} or from fast rotating neutron stars (millisecond pulsars) \cite{Bogdanov:2019qjb}.

Finally, it is important to mention a further source of experimental data that has been developed during the last years, and which can provide a lot of information about neutron stars. The detection of gravitational waves that resulted from the merging of binary neutron star systems has allowed to understand the origin of the heavy elements of the periodic table, but also permits to extract information about the equation of state of nuclear matter at high densities \cite{agathos2015constraining}, by analysing the gravitational wave spectra. These observations are still not sufficiently accurate to completely determine the nuclear matter EoS at high densities, but the growth of observational data and the improvement of the techniques to study them will lead to important progress in this field in the near future. 

%%%%%%%%%%%%%%%%%%%%%%%%%%%%%%%%%%%%%%
\section*{Acknowledgements}
%%%%%%%%%%%%%%%%%%%%%%%%%%%%%%%%%%%%%%
The authors acknowledge financial support from the Ministry of Education, Culture, and Sports, Spain (Grant No. FPA2017-83814-P), the Xunta de Galicia (Grant No. INCITE09.296.035PR and Conselleria de Educacion), the Spanish Consolider-Ingenio 2010 Programme CPAN (CSD2007-00042), Maria de Maetzu Unit of Excellence MDM-2016-0692, and FEDER.

\bibliographystyle{ieeetr}
\bibliography{refs}

\end{document}